\documentclass[sn-mathphys-ay]{sn-jnl}

\usepackage{url} 
\usepackage{lineno}
\usepackage[inline]{trackchanges} 
\usepackage{amsmath}
\usepackage{commath}
\usepackage{amssymb}
\usepackage{amsbsy}
\usepackage{natbib}

\usepackage[utf8]{inputenc}
\usepackage{nomencl}
\usepackage{bbold}

\graphicspath{ {./images/}}

\usepackage{graphicx}%
\usepackage{multirow}%
\usepackage{amsmath,amssymb,amsfonts}%
\usepackage{amsthm}%
\usepackage{mathrsfs}%
\usepackage[title]{appendix}%
\usepackage{xcolor}%
\usepackage{textcomp}%
\usepackage{manyfoot}%
\usepackage{booktabs}%
\usepackage{algorithm}%
\usepackage{algorithmicx}%
\usepackage{algpseudocode}%
\usepackage{listings}%

\begin{document}

\title[Article Title]{Pore-space partitioning in geological porous media using the curvature of the distance map}


\author*[1,2]{\fnm{Ilan} \sur{Ben-Noah}}\email{ilanb@volcani.agri.gov.il}

\author[2]{\fnm{Juan j.} \sur{Hidalgo}}\email{juanj.hidalgo@idaea.csic.es}

\author[2]{\fnm{Marco} \sur{Dentz}}\email{marco.dentz@csic.es}

\equalcont{All authors contributed to the study conception and design. Material preparation, data collection and analysis were performed by Ilan Ben-Noah. The first draft of the manuscript was written by Ilan Ben-Noah and all authors commented on previous versions of the manuscript. All authors read and approved the final manuscript.}

\affil*[1]{\orgdiv{Department of Environmental Physics and Irrigation}, \orgname{Institute of Soil, Water and Environmental Sciences, The Volcani Institute, Agricultural Research Organization}, \orgaddress{\city{Rishon LeZion}, \country{Israel}}}

\affil[2]{\orgdiv{Institute of Environmental Assessment and Water Research (IDAEA)}, \orgname{Spanish National Research Council (CSIC)}, \orgaddress{\city{Barcelona}, \country{Spain}}}


\abstract{Media classification and the construction of pore network models from binary images of porous media hinges on accurately characterizing the pore space. 
We present a new method for (i) locating critical points, that is, pore body and throat centers, and (ii) partitioning of the pore space using information on the curvature of the distance map (DM) of the binary image. Specifically, we use the local maxima and minima of the determinant map of the Hessian matrix of the DM to locate the center of pore bodies and throats. The locating step provides structural information on the pore system, such as pore body and throat size distributions and the mean coordination number. The partitioning step is based on the eigenvalues of the Hessian, rather than the DM, to characterize the pore space using either watershed or medial axis transforms. This strategy eliminates the common problem of saddle-induced over-partitioning shared by all traditional marker-based watershed methods and represents an alternative method to determine the skeleton of the pore space without the need for morphological reconstruction.

\subsubsection*{Plain Language Summary}
Recent and ongoing developments in small-scale and 3D imaging capabilities and the manufacturing technology of microfluidic devices enable experimental visualization and analysis of the micro-scale (microns) pore system geometry and connectivity. These developments drive fluent research on the mechanisms of pore-scale flow and transport. Moreover, these advancements also allow better characterization and extraction of more realistic pore network models. 
Which, in turn, are commonly used to evaluate flow and transport phenomena in geological and industrial porous media. Here, we suggest a new approach to process binary images (i.e., a description of the solid and pores as zeros and ones) by using a transformed map that holds information about the image topography to locate the critical points of the pore system and for the partitioning of the pore space to distinct pores. This method yields some key features that can be used to construct discrete pore networks and derive more detailed information on the pore system geometry.
}




\maketitle

\section{Introduction}\label{sec1}

Pore network models have been extensively used to simulate flow and transport processes through porous media~\citep{blunt2017}. The performance of these models hinges on the accurate description of the pore space \citep{sahimi2011, zhao2020, Mehmani2017, ben2024network}. The last two decades have seen enormous progress in the imaging and characterizing of the micro-scale structure and multi-phase flow and transport processes in porous media \citep{wildenschild2013,bultreys2016imaging}. These new abilities are accompanied by a growing number of partitioning algorithms to characterize the geometrical and topological features of the pore system \citep{warner1989, lee1994, pauli1997, lindquist2000, oren2002, rabbani2014, thompson2008, brun2010, ngom2011, safari2021}. In this paper, we propose a method for pore network characterization that uses curvature information of the distance map for locating critical points and partitioning the pore space.  

Pore network extraction from a gray-scale image of a porous medium sample can be split into two fundamental steps. First, the image is converted into voxels belonging to the void and solid phases. Methods for image binarization are reviewed, for example, in the papers by \cite{wildenschild2013} and \cite{bultreys2016imaging}. Second, the resulting image is processed to extract the features of interest (e.g., pore bodies and throats) and their geometrical properties (e.g., hydraulic radius, length, volume, and shape descriptors). We focus here on the second step, i.e., we assume that we have binary images that distinguish between void and solid phases. Feature identification can be further split into two stages, i.e., the location of critical points (pore bodies and throats) and image partitioning.

The location of the critical points is mostly done based on three methods: the distance map, maximal balls, and medial axis transforms. The first one defines the critical points as the maxima and saddle points of the distance map (DM) or distance transform. The DM is the Euclidean distance of each point in the void phase from its nearest void-solid phases interface \citep{borgefors1984distance}. 
The significance of the critical points draws from their different roles regarding the hydraulic properties of a medium. The pore body size, denoted by the maxima of the DM, relate to retention and is reciprocal to the driving force for capillary flows (of a wetting phase), while the throat size, given by the saddle points of the DM, relates to the resistance to the flow \citep{sahimi2011}.
While locating the maxima of the DM is relatively straightforward, locating the saddle points is more challenging, even for regular media, and requires additional image transformation, as discussed further below. For complex porous media, the discretization of the pore space depends on the image resolution, which may create "false" critical points. These mathematically correct "false" critical points may be the results of the image resolution, numerical inaccuracies, or subjective morphological definition of "real" critical points.  Moreover, plateaus in the DM (either real or resolution-limited) can result in errors in the classification and spatial location of the critical points \citep{gostick2017}.  

The maximal balls method \citep{dong2009} inscribes spheres \citep{silin2006} at each point of the void space such that they touch the solid boundaries (equivalently, disks in two-dimensional images, \cite{solomon2011fundamentals}). A sphere is called a maximal ball if it is not contained in another sphere. The centers and radii of the maximal balls represent the locations and sizes of the pore bodies. The minimal balls between pores represent the locations of throat centers and throat radii. This method is computationally expensive. 

The medial axis method first erodes (or burns) the pore system to its skeleton. To illustrate the medial axis, it is instructive to consider a fire that consumes the pore space from its boundaries at a constant speed. The line at which the burning fronts meet is the medial axis or skeleton of the pore space. The number of layers burned, termed the {\em burn number}, provides information on the shortest distance to the solid-void interface at each skeleton point \citep{baldwin1996}. The critical points are defined by the local extrema of the burn numbers along the skeleton. The pore body centers are located on the maxima, and pore throats on the minima \citep{liang2000}. Less accurately, the pore centers can be located at the junctions (vertices) of the skeleton \citep{lindquist1996}. However, this method overlooks pore bodies and throat centers located on plateaus in the burn number that are commonly found in disordered media.

The information obtained from the locating process (e.g., size distribution and mean coordination number) is a fundamental pore system characteristic required for constructing simple pore network models \citep{sahimi2011}. For a detailed network representation, which accounts for the distribution of coordination numbers, pore shapes, and pore volumes, a partitioning of the void space into distinct pores is required. The critical point locating process plays a central role in pore space partitioning. We focus here on three image partitioning methods that have been used for pore network extraction, medial axis partitioning \citep{baldwin1996, lindquist1999, jiang2007, lee1994, liang2000, youssef2007}, watershed partitioning \citep{meyer1994, gostick2017, soille1990}, and Delaunay tessellation \citep{mason}.    

\paragraph*{Medial axis.} Medial axis partitioning is a two-stage topology-centered method that uses directly the binary image. The first step is locating the critical points, which have been described above. After this step, the pore space is partitioned into its constituent pores by identifying pore bodies as clusters of void voxels that are bounded by solid voxels and pore necks. The latter is defined as the cross-sectional surfaces located at the throat centers \citep{liang2000}. This is a challenging step because it is subject to over-partitioning due to local minima on the skeleton that are not actual throats \citep{liang2000} and because vertices of the skeleton junction are not necessarily pore centers \citep{blunt2017}.

\paragraph*{Watershed.} The marker-based or {\em seeded} watershed partitioning method, termed here {\em DM-based watershed}, is a morphology-centered method. It is convenient to illustrate this method in terms of the reversed DM, so that the low plains are at the grain centers. These minima points (pore body centers) are termed seeds or markers. Then the algorithm {\em floods} the reverse map with {\em water} sourced at these locations (markers) so that it marks all the neighboring sites (pixels) that are of the same or smaller value. The algorithm keeps raising the water level until it reaches the local maxima of the reverse map, which correspond to the minima or saddles of the DM. Here, we use the DM-based watershed as the reference classical partitioning method because it is commonly used and easy to implement. Typically, the main intrinsic limitation of this method is the over-partitioning caused by possible false interpretations of saddles as markers. This issue can be addressed by using a complementary search algorithm \citep{gostick2017}, which is time-consuming and computationally expensive. 

\paragraph*{Delaunay tessellation.}
Delaunay tessellation is a common method to subdivide (partition) the void space of regular or random sphere packing. This method forms tetrahedrons (triangles in 2D) connecting adjacent grain centers using Delaunay triangulation. This is done in such a way that, for a given set of discrete points, no point is inside the circumcircle of any tetrahedron in the triangulation (i.e., there is no overlap). The distinct pores are defined by the voids inside the tetrahedrons \citep{mason} and the throats by the void area on the sides of the tetrahedron \citep{mellor, bryant}. While this method proves useful for extracting pore geometry when the set of points corresponds to locations of nearly spherical well-sorted grains \citep{smith1987mercury}, it is less useful for angular, large aspect ratio grains. Moreover, this method does not account for the morphology of the pore space. Furthermore, it is limited to a uniform coordination number distribution of four (for 3D), and its subdivision of pores can lead to different identification of pore location and sizes compared to morphology-centered methods \citep{alRaoush}.   

In this paper, we present a framework based on the curvature map of the DM for (i) locating the critical points and (ii) for pore space partitioning. Geometrically, the DM is a topographic map, with peaks at the pore centers, ridges connecting the pores, and valleys connecting the solid grains. From this point of view, it is clear that the slope (gradient) and curvature give information about the morphology and topology of the pore space. Quantifiers of the curvature (i.e., the divergence of the normalized gradient of the signed distance function) are given by the Hessian matrix, its determinant, and its eigenvalues. Here, we use the determinant of the Hessian matrix to locate the ridges of the DM, that is, the throat centers. The peaks of the DM determine the pore body centers. We propose two novel methods for image partitioning based on the watershed transform and medial axis, respectively. Unlike classical approaches, we apply these transforms to the map of curvature and the maps of the eigenvalues of the Hessian matrix. The proposed approach is similar in spirit to the Hessian affine region detector in computer vision \citep{mikolajczyk2002} that is based on the Hessian of the intensity map of an image to identify points of interest. \cite{chen2008} used the gradient of the DM to identify points of interest and the determinant of the Hessian to identify saddle points for granules partitioning of 2D binary images. The relation between the eigenvalues of the Hessian and the location of pore throats was pointed out by \cite{blunt2017}. However, to the best of our knowledge, this relation has not been used to analyze the pore space of 2D or 3D porous media nor to locate its critical points. 

The paper is organized as follows. Section~\ref{sec:algorithm} presents the proposed algorithm for critical point location and pore space partitioning. It first presents the distance transform and filtering steps to obtain the determinant of the Hessian matrix and its eigenvalues. Then, it describes the method for locating critical points and pore space partitioning on the curvature map. Section~\ref{sec:app} demonstrates and validates the proposed algorithms for porous media of different complexity.  

\section{Algorithm for critical point location and pore space partitioning\label{sec:algorithm}}

In this section, we present the algorithm to determine the location of pore throats and bodies and two methods based on the medial axis and watershed transforms to partition the pore space. The algorithm is based on identifying critical points of the DM \citep{borgefors1984distance}. The pore centers are located at the maxima, and the throat centers are at the saddle points of the DM. These definitions are invariant under translation, uniform magnification, and rotation in spatially variable transformations \citep{eberly1994}. Unlike other methods, the proposed algorithm uses not only the information on the peaks of the DM but also its curvature contained in its Hessian matrix. The maxima and saddle points are located using the determinant of the Hessian matrix. The two new methods for image partitioning are based on the eigenvalue map ($\lambda$-map) of the Hessian matrix. The partitioning methods used belong to the watershed and medial-axis families. 

\begin{figure}
\centering
 \includegraphics[width=\textwidth]{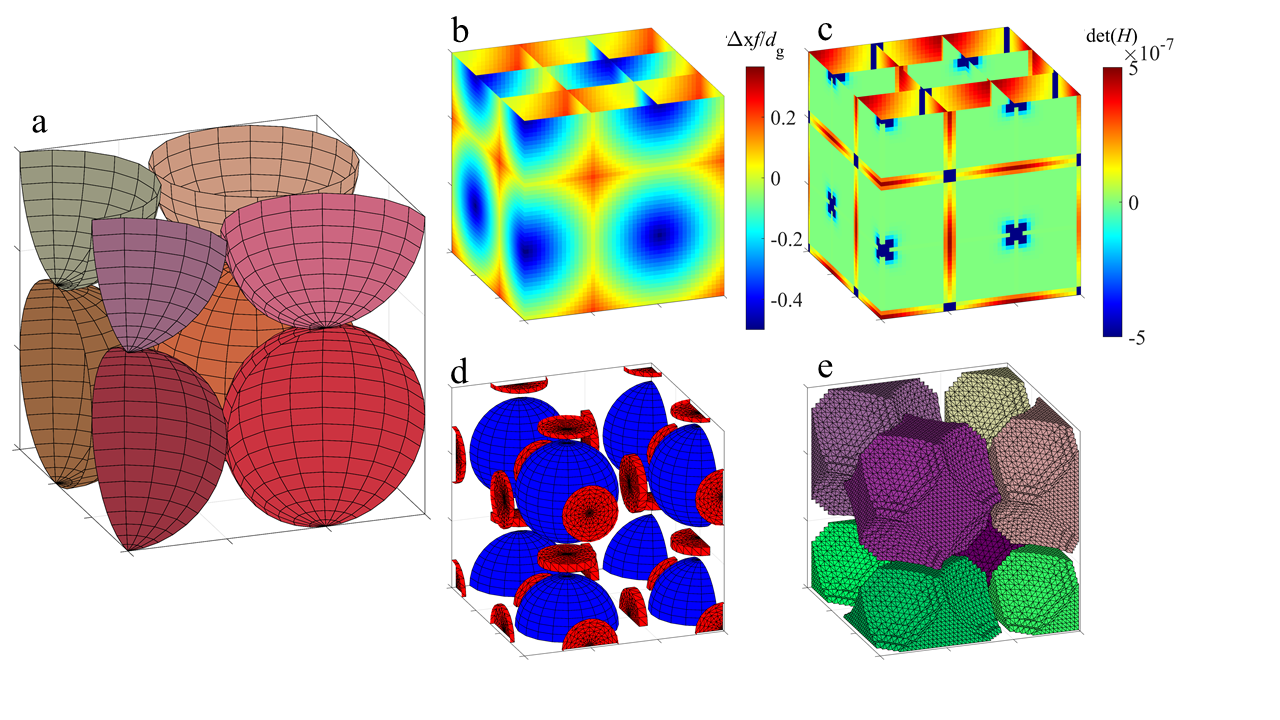}
 \caption{Illustration of the locating and partitioning processes for cubic packing of spheres. a) image of the cubic spherical-grains packing, b) distance map (DM), c) determinant of the Hessian matrix of the distance transform, d) pore centers and radii ("maximal spheres", blue) and throat centers, radii, and directions (red cylinders), and e) partitioned pores with the $\lambda$-based watershed method.} \label{fig:cubic} 
 \end{figure}

The three fundamental steps of the algorithm are illustrated in Figure~\ref{fig:cubic} for a 3D cubic packing of uni-sized spherical grains. They consist in:
\begin{enumerate}
    \item {\em Distance transform and curvature maps:}  The binary image shown in Figure~\ref{fig:cubic}a is used to calculate the DM, which is shown in Figure~\ref{fig:cubic}b. Grains are located in the negative values, while the pores are in the positive values of the DM. Then, the Hessian matrix and its determinant map shown in Figure~\ref{fig:cubic}c are calculated from the numerical derivatives of the DM. If necessary, filters can be applied to the DM to reduce the identification of spurious critical points.
    
    \item {\em Locating critical points:} The local maxima of the DM (red color in Figure~\ref{fig:cubic}b) are used to locate the pore body centers (blue spheres in Figure~\ref{fig:cubic}d). The local minima of the determinant map of the Hessian (blue color in Figure~\ref{fig:cubic}c) are used to locate the throat centers (red cylinders in Figure~\ref{fig:cubic}d). Artificial critical points are removed by filtering the DM and by applying a complementary critical point dilution, as discussed in detail in Section \ref{sec:Complementary}.
    
    \item {\em Pore space partitioning:} Lastly, the partitioning of the pore space is done based on the map of the eigenvalues ($\lambda$) map of the Hessian. This is done via the novel methods presented in this work, i.e., $\lambda$-based watershed (Figure~\ref{fig:cubic}e) and $\lambda$-based medial axis (not shown). 
\end{enumerate}

In the following, we describe these steps in detail. 

\subsection{Step 1. Distance transform and curvature maps}
The distance map is defined here by the Euclidean distance between each pixel and the nearest interface between the pore domain $\mathbb P$ and the matrix domain $\mathbb S$ of all locations in the solid. The set of points in the solid that are located at the interface between the solid and the pore space is denoted by $\partial \mathbb S$. Thus, the distance map can be written as
\begin{linenomath}
\begin{equation} \label{eq:dist.map}
f(\mathbf x_k) = \frac{\text{sign}[\mathbb{I}(\mathbf x_k \in \mathbb P)-0.5]}{\Delta x}\min_{\mathbf x_j \in \partial \mathbb S}\left(\lVert \mathbf{\mathbf x_k - \mathbf x_j} \rVert \right),
\end{equation}
\end{linenomath}
where $\Delta x$ is the Pixel size, $\lVert \mathbf{\mathbf \cdot} \rVert $ denotes the Euclidean norm, $\mathbf x_k$ is the location of voxel $k$, the signum function $\text{sign}[\cdot]$ equals $1$ for positive and $-1$ for negative values, $\mathbb I(\cdot)$ is the indicator function, which is one if its argument is true and zero else.  The signum function $\text{sign}[\cdot]$ is used to assign negative values to the distance map $f(\mathbf x)$ for locations $\mathbf x$ within the solid matrix. Assigning negative values within the complementary-to-the-pore matrix \citep{kaestner2008} reduces the number of irrelevant inflection points at the grain surfaces and the truncating error of the numerical gradient operator, described in the following sub-section. For the distance transform in irregular media, we use the MATLAB ~\citep{MATLAB} function {\em bwdist}, which extracts the Euclidean distance map from binary images. One should bear in mind that the distance map from a binary image introduces a truncation error for curved objects at the scale of the resolution.

The curvature map is described by the Hessian matrix,
\begin{linenomath}
\begin{equation} \label{eq:Hessian}
H_{ij} = \frac{\partial^2 f(\mathbf x)}{\partial x_i \partial x_j}. 
\end{equation}
\end{linenomath}
Thus, the Hessian holds information about the DM landscape. The derivatives of the DM are calculated using a first-order central difference scheme for the interior data points. First-order forward or backward difference schemes are used for the edge points. The Hessian matrix is symmetric, and therefore its eigenvalues are real. Moreover, these methods are isotropic, i.e., independent of the alignment of the axis on the direction of the media. The eigenvalues $\lambda_i(\mathbf x)$ are defined at each point in space by the roots of
\begin{linenomath}
\begin{align}
    \det\left[\mathbf H(\mathbf x) - \lambda_i(\mathbf x) \mathbb{1} \right] = 0,
\end{align}
\end{linenomath}
where $\mathbb{1}$ is the identity matrix. The index $i = 1,\dots,d$, where $d$ denotes the dimensions of space. In the following, we denote by $\lambda_1(\mathbf x)$ the smallest of the $d$ eigenvalues at each point.

It should be noted that the determinant of the Hessian describes the Gaussian curvature, and the eigenvalues of the Hessian are the principal curvatures of the DM. The eigenvalues $\lambda_i$ of the Hessian matrix at each point in space quantify the curvature of the field gradient in various directions. A small eigenvalue indicates a small curvature (low rate of change) in the eigen-direction \citep{aragon2007}. Thus, the eigenvalues of the Hessian matrix reflect the ridges and valleys of the DM. 
At the saddle points, $\lambda_1<0$ and $\lambda_3>0$, while the sign of $\lambda_2$ indicates the index of the saddle, i.e., the number of descending directions of the saddle \citep{armstrong}. 
For the case of throats located at the center of the minimal surface between three solid grains, $\lambda_2$ is positive (an index-1 saddle). 
This information can be used as a screening criterion for false saddle points.

\subsection{Step 2. Identification of critical points in the distance map}
This section presents a method to locate the critical points by combining the DM~\eqref{eq:dist.map} and the Hessian matrix~\eqref{eq:Hessian}. Throat centers are located on the saddle points of the DM, and grain and pore body centers are located on the local maxima. The Hessian matrix is used to derive information about the distance map landscape. At saddle points, the determinant of the Hessian is negative, and its eigenvalues are of opposite sign. A strong negative value of the determinant indicates the presence of multiple edges. Thus, \cite{lakemond2011} concluded that the local minima of the determinant of the Hessian can be used to identify the location of saddle points of the DM.

\subsubsection{Identification of local extrema of the DM}

The method first identifies the local extrema of the DM. A pixel with a DM value larger or smaller than all its surrounding pixels, based on $8-$connected elements for 2D and $26-$connected for 3D, is considered a local extreme. Then, the classification is done using the Hessian matrix. The grain and body centers are located on the local maxima with positive values of the Hessian determinant map (Figure~\ref{fig:cubic}c). The binary image can be used to directly exclude all the points that belong to grains ($\mathbb S$). The radii of the pore bodies are given by the values of the distance map at their center location.

\subsubsection{Identification of saddle points of the DM}
The throats are the surfaces that constitute the local minima cross-sections. The saddle points are located at the farthest points from the solid matrix on these surfaces (obtaining a maximal value on the DM).
The saddle points of the distance map are identified as the local negative minima of the determinant map \citep{lakemond2011} (Figure~\ref{fig:cubic}c). The radii of the pore throats are given by the values of the distance map at the saddle point locations. The orientation of the throat (red cylinders,  Fig.~\ref{fig:cubic}d) can be evaluated from the maximal gradient of the DM at the saddle point \citep{chen2008}. Alternatively, the throat orientation can be found by the normal vector to the line (for 2D, circle for 3D) drawn between the two (three for 3D) closest solid-pore interfaces. These solid-pore intersects are located at distances $f(\mathbf x _s)$ from the center of the throat, where $\mathbf x_s$ are the coordinates of the pore throats' centers. The variability of pore throat orientations can give information about the tortuosity of the medium.

\begin{figure}
\centering
 \includegraphics[width=\textwidth]{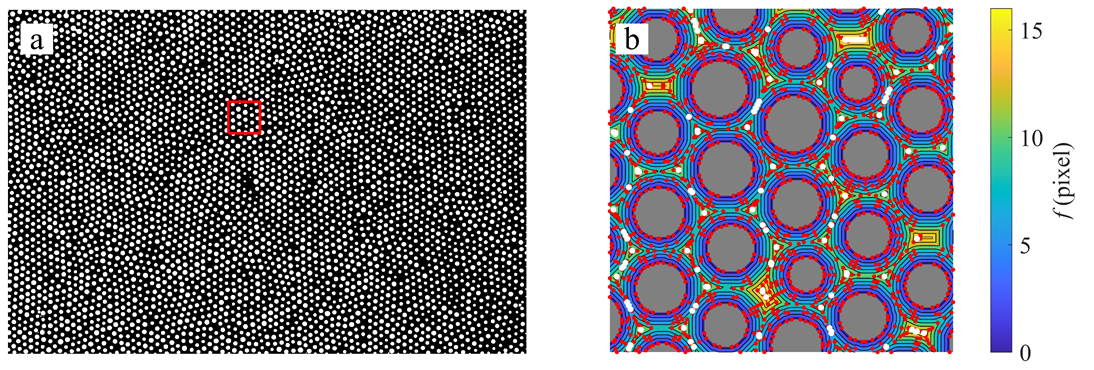}
  \caption{a) Binary image of a two-dimensional disordered medium consisting of randomly placed circular disks of variable size. The red square depicts the zoomed-in area of b) the distance transform map ($f$, zoomed in) and the critical point locations (without filtering). Grains are in gray, pore centers are white, and the throat centers' are red points.} \label{fig:sat_2D_unfiltered} 
 \end{figure}

\subsubsection{Removing artificial critical points in the distance map} \label{sec:Complementary}

For regular media, the DM and, in turn, the determinant of the Hessian matrix is smooth enough for the critical points to be directly evaluated. For more complex disordered media, identifying the critical points is more difficult because of the binary images' finite resolution and the pore space's complex geometries with different scales. This may lead to identifying false critical points on plateaus or ridges of the distance or determinant maps. Typically, the plateaus formed in the DM are bound to create additional false points rather than ignoring real ones \citep{gostick2017}. This is illustrated in Figure \ref{fig:sat_2D_unfiltered}, which shows a two-dimensional disordered granular medium. For this medium, the unfiltered distance map includes many false local peaks (white points in Fig.~\ref{fig:sat_2D_unfiltered}b) and saddle points (red points).

The identification of false critical points can be mitigated by a filtering process of the DM. In this work, we use single-layer Gaussian and pyramid Gaussian filters to reduce the possibility of false positives in the critical points location step. The major merit of the Gaussian filter is that it does not create new artificial extrema. A description of the filters can be found in appendix~\ref{app:filtering}.
Although filtering can help improve the accuracy of critical point location, it may not be efficient enough in some situations, e.g., when the number of critical points between iterations changes slightly. Therefore, after the filtering step eliminates most of the false critical points, a complementary scheme can be implemented on the remaining critical points to reduce the number of false positives further. We propose a complementary critical point dilution method based on the following three-stage elimination criterion:
\begin{enumerate}
    \item  overlapping pore bodies are eliminated, keeping the largest pore (consistent with the maximal ball representation).
    \item  overlapping saddle points are eliminated, keeping the smallest values, 
    \item  saddle points whose circle (or sphere) is completely embedded within a pore body circle (or sphere) are eliminated.
\end{enumerate}
A critical point is considered to overlap with an adjacent one when it is located within the maximal circle (or sphere) of the adjacent critical point (i.e., pore bodies or throats).
 The efficiency of the filtering compared to the complementary scheme is case-specific and depends on the size of the image, the number of false points, and the filtering parameters.

\subsection{Step 3. Image partitioning into pore regions}

A detailed representation of the pore network requires partitioning the pore space. This section presents two methods to partition the void space into discrete pores. The first method belongs to the medial axis family, and the second to the watershed family. Instead of using the DM, these methods use the information about its curvature contained in the Hessian matrix, that is, the eigenvalues ($\lambda_i$) of the Hessian matrix, which reflect the ridges and valleys of the DM. 

\subsubsection{$\lambda$-based medial axis partitioning} \label{sec:lma} 

\begin{figure}
\centering
 \includegraphics[width=\textwidth]{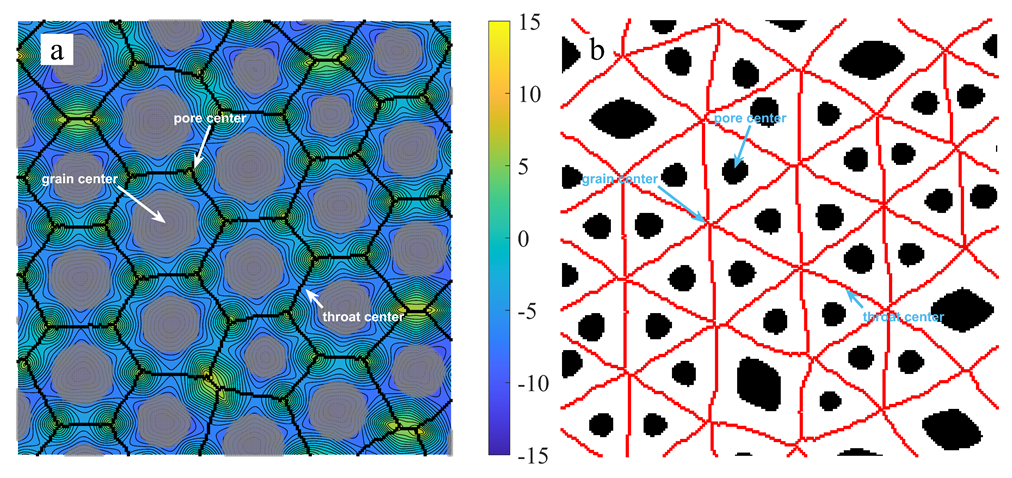}
 \caption{a) Eigenvalues $\lambda_1$ map of the Hessian matrix for the saturated 2D medium (Figure~\ref{fig:sat_2D_unfiltered}a), where the black lines represent the skeleton of the original binary image, grains are depicted in gray (slightly transparent). b) Binary map of $M_{\lambda_1}$ \ref{eq:lambda_1_binary}. were the white regions represent the $\lambda_1<0$ values and black regions the $\lambda_1>0$, red lines presents the skeleton of the white regions. For these analyses, the Gaussian pyramid filtering parameters of $\gamma=1$, $\sigma_{0}=3$, and $\Delta \sigma=0.01$ were used as described in ~\ref{app:filtering}.} \label{fig:lambdas_2Dsat}  
 \end{figure}

The pore space medial axis is formed by the geographic ridges of the DM. As outlined in the Introduction, the medial axis is commonly evaluated using a morphological thinning process \citep{baldwin1996}. A binary matrix $M_{\lambda_1}$ is derived from the $\lambda_1$ map such that all values smaller than $0$ are assigned $1$ and $0$ else. That is,
\begin{linenomath}
\begin{equation}\label{eq:lambda_1_binary}
M_{\lambda_1}(\mathbf x)=\mathbb I(\lambda_1(\mathbf x)<0) 
\end{equation}
\end{linenomath}
Figure~\ref{fig:lambdas_2Dsat}a depicts the skeleton (black lines) of the binary image in Figure~\ref{fig:sat_2D_unfiltered}a and the map (contours) of the smallest eigenvalues ($\lambda_1$) of the Hessian of the DM. Fig.~\ref{fig:lambdas_2Dsat}b) shows the binary matrix $M_{\lambda_1}$. The negative values of the smallest eigenvalue $\lambda_1$ (white regions depict the valleys of the DM so that its skeleton, marked by red lines, connects the grain centers colored in gray). The skeletonization (i.e., the erosion of a binary image to its skeleton) of $M_{\lambda_1}$ forms a line that passes through the pore throats (red lines in Figure~\ref{fig:lambdas_2Dsat}b). The union of this skeleton and the complementary phase (i.e., the grains) encloses and partitions the different pores. This process allows for a straightforward pore partitioning to evaluate the detailed characteristics of the pore system. However, partitioning based on the $\lambda_1 < 0$ binary image sometimes unites pores that are connected with large throats (Fig.~\ref{fig:watershed}a). This issue can be resolved by using a threshold smaller than $0$ for the binarization, keeping in mind that using too small a threshold will cause the splitting of large pores. Moreover, this partitioning method seems to miss some extremely small pores (e.g., left lower corner of Fig.~\ref{fig:watershed}a). Despite these limitations, this partitioning method seems to be a good compromise, given its straightforwardness and efficiency. The applicability of this compromise depends on its use. In addition, the intersections of ridges (black lines in  Figure~\ref{fig:lambdas_2Dsat}a)  and valleys (red lines in  Figure~\ref{fig:lambdas_2Dsat}b) denote saddle points, which mark the pore throat centers. This can be used as an alternative method to locate pore throats. 

\subsubsection{$\lambda$-based watershed partitioning} \label{sec:lws}

\begin{figure} [h]
\centering
 \includegraphics[width=\textwidth]{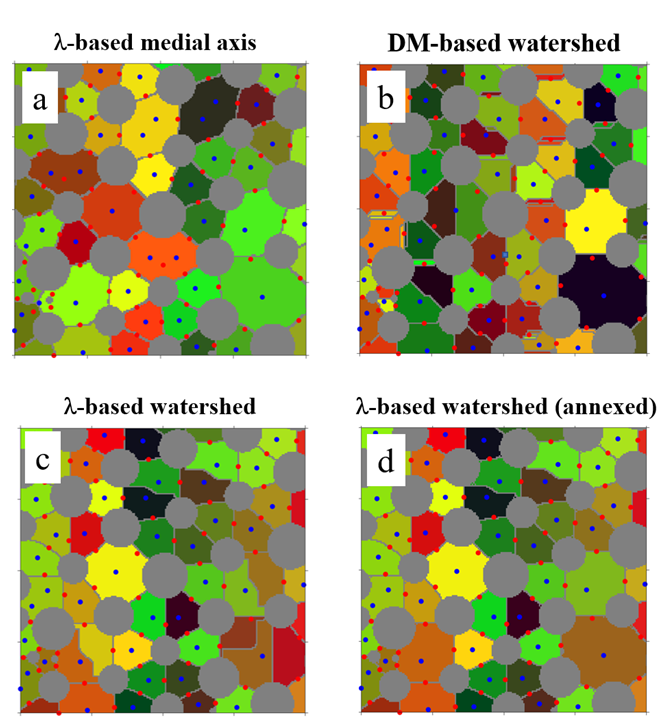}
 \caption{Partitioned pores of the saturated $2D$ using a) the skeleton of $M$ \eqref{eq:lambda_1_binary} and the watershed method \protect\cite{meyer1994} on b) the distance map, c) on the $\lambda_1$ (Fig.~\ref{fig:lambdas_2Dsat}b) map, and d) after the annexation of pores without a critical point. Different pores have different colors. Blue points represent the located pore bodies' maxima (peaks), and red points represent the saddle points. The solid grains are gray. For a color interpretation of this image, we refer readers to the online version.}\label{fig:watershed}  
 \end{figure}

Watershed partitioning is sensitive to the distinction between saddles and extreme points \citep{meyer1994, gostick2017, soille1990}.  When using the watershed transform directly on the DM, many false small pores are centered around the saddle points (Figure~\ref{fig:watershed}b). In order to mitigate this problem, we use the marker-based watershed algorithm described in \cite{meyer1994}, but instead of applying it to the distance map, we transform the $\lambda _1$ map (Figure~\ref{fig:lambdas_2Dsat}a), which gives the topographic valleys of the DM. 
The extrema of $\lambda_1(\mathbf x)$ coincide with the critical points of the DM so that the pore centers are at the maximal values and the grain centers are at the minima. Similarly, the pore throats are at the saddle points in both the DM and $\lambda _1$ map. 
However, in the $\lambda_1(\mathbf x)$ map, the saddle points obtain negative values (as the grains), while in the DM, they are positive (as the pores). As further discussed below in Section~\ref{sec:app}, the main benefit of using the $\lambda _1$ map instead of the DM is that the saddle points are not mistakenly assigned as markers for the watershed. This resolves the issue of false pores around the saddle points (Figure~\ref{fig:watershed}c). However, the new method may create false pores by splitting up large ones. To correct this, we use a search algorithm to find pores identified by the watershed algorithm that does not contain a body center. Then, we annex pores without a maximal point (blue point in Fig.~\ref{fig:watershed}) with the pore with the nearest pore center (Figure~\ref{fig:watershed}d).

\section{Validation and demonstration\label{sec:app}}

Pore partitioning is a challenging task because porous materials (especially naturally occurring ones) do not necessarily comply with our conceptual view of breaking the space into pores and throats \citep{Mehmani2020}. It depends on the type of media, the quality of the input data, and also on the application. Thus, it is difficult to identify or qualify a single best method for pore space partitioning. This notion emphasizes the importance and value of alternative methods and approaches and, at the same time, reflects on the difficulty of comparing and evaluating different methods. Thus, in this section, we validate the proposed method against analytical results for simple granular media, and by comparison with the classical DM-watershed and Delaunay Tesselation methods. Furthermore, we demonstrate the applicability of the proposed methods to complex porous media.  

Specifically, in Subsection~\ref{sec:SC}, we validate the proposed methods against analytical solutions of pore and throat locations and sizes. These solutions are restricted to simple granular media such as simple cubic (SC) or body-centered cubic (BCC) alignments of spheres. We also use, in subsection \ref{sec:Finney}, a random packing of spheres \citep{Finney} to compare the proposed partitioning methods to the commonly used Delaunay tessellation (DT) method. In Subsection~\ref{sec:2d}, we use images from two-phase flow experiments in a manufactured two-dimensional (2D) millifluidic device \citep{jm2017} to evaluate the filtering process and the sensitivity of the proposed methods to image resolution compared to the traditional DM-watershed partitioning method. Finally, in Subsection~\ref{sec:berea}, we use a Berea sandstone \citep{berea} rock sample to compare the proposed $\lambda$-based medial-axis method to the DM-watershed method and demonstrate the applicability of the algorithm to images from CT scans of a Bentheimer sandstone \citep{berea} and a tight Carbonate \citep{Carbonate}.

\subsection{Three-dimensional sphere packings\label{sec:SC}} 
Here, we consider three types of sphere packings. A simple-cubic (SC) and a body-centered cubic (BCC) packing, where the sphere centers are located at the vertices of a cubic lattice (Figure~\ref{fig:cubic}a). The grain diameter $d_g$ is chosen such that neighboring grains touch. The cubic packings are used as a benchmark of the method on account of having a clear, visible, and perceivable pore body and throat system. The cubic sphere packings are simulated in a volume that contains five grains in each direction and with a resolution of $\Delta x = \frac{d_g}{32}$ for the SC and $\Delta x = \frac{d_g}{16\sqrt{3}}$ for the BCC packing. The body-centered cubic (BCC) packing can be found in the supplementary material. Furthermore, we consider a Finney medium, a random packing of spheres of different sizes \citep{Finney} shown in Figure~\ref{fig:Finney}a. 

\subsubsection*{Step 1. Distance and curvature maps} 
The DM defined by Eq. \eqref{eq:dist.map} can be evaluated as the minimum difference between the Euclidean distance of each pixel to each of the grain centers and their radii. This evaluation of the distance transform is not subject to numerical truncation and is expected to create fewer false critical points than the numerical evaluation of the Euclidean distance, which is based on the distance of each pixel from the nearest pixel on the solid-pore interface. Here, the locations and sizes of the grains in the regular packings are known. For the Finney packing, the central locations of grains are evaluated as the centroid (the center of mass) of each grain. Distinct grains are distinguished using the pixel-based distance map. Notice that this method may falsely locate the center of the grains at the image borders (for cutoff grains).  

\subsubsection*{Step 2. Locating critical points} 
For both the SC (and BCC) arrays of regular spherical grains, the locating method works remarkably well (when using the exact DM described above) without any need for filtering or complementary dilution of critical points. Both the number and location of the pore bodies and throat are exact, as shown in figures \ref{fig:cubic} and the supplementary material (Figure SF1). 
The SC packing has a porosity of $1-\frac{\pi}{6}\approx 0.476$. Moreover, this medium has a constant pore body diameter (blue spheres in Figure~\ref{fig:cubic}d) of $0.73d_{g}$ and a uniform pore throat diameter of $0.41d_{g}$, which are in agreement with the theoretical values of $(\sqrt{3}-1)d_{g}$ and $(\sqrt{2}-1)d_{g}$, respectively. Notice that here, we define the pore body ($d_b$) and throat ($d_t$) diameters by the maximal sphere and circle that can fit within the pore element. Also, this medium has a constant coordination number (i.e., the mean number of interconnected pores) of $Z=6$ (Fig.~\ref{fig:cubic}d). The BCC array has a porosity of $1-\frac{\sqrt{3}\pi}{8}\approx 0.320$, and constant pore body and throat diameters in agreement with the theoretical values of pore diameter ($(\sqrt{\frac{5}{3}}-1)d_g$) and throat diameter ($(\sqrt{\frac{3}{8}}-\frac{1}{2})d_g$). The coordination number is $Z=4$; see the supplementary material.
\subsubsection*{Step 3a. $\lambda$-based watershed partitioning--regular packings}
The pore body, in terms of the maximal enclosed sphere (denoted by the blue spheres in Figs.\ref{fig:cubic}d), obviously underestimates the actual pore volume estimated by partitioning of the void (Figs.\ref{fig:cubic}e). The actual pore volume can be evaluated by counting the voxels of each pore in the partitioned map. The throat areas can be evaluated from the number of voxels that border two different pores.  
The actual pore volume, evaluated by the $\lambda$-based watershed partitioning of these media, is similar to the theoretical values of $(1-\frac{\pi}{6})d_{g}^3$ and $(\frac{8-\pi\sqrt{3}}{36\sqrt{3}})d_{g}^3$ for the SC and BCC packings.
For the regular media, the $\lambda$-based medial axis partitioning gives exactly the same results as the $\lambda$-based watershed partitioning, and it is not presented in this subsection to avoid repetition. Similarly, it should be mentioned that the DM-based watershed gives similar results for these media. From this, we conclude that i) the new partitioning methods work, and ii) regular media provide a bottom line for any partitioning method, but are limited as test cases to evaluate different partitioning methods.

%
The definition of the pore throat as the maximal disk that fits the throat (denoted by the red cylinders in Fig.\ref{fig:cubic}d) underestimates the actual throat surface area. For example, according to the theoretical value, also obtained by this locating method (calculated from its diameter), the circle area of the pore throat for the SC is $\frac{\pi(\sqrt{2}-1)^2d_g^2}{4}$ which stands for about $63\%$ of the actual throat surface area ($(1-\frac{\pi}{4})d_{g}^2$). This difference might reflect on the applicability of using the maximal disk definition of the pore throat, rather than the exact surface area and geometry, to evaluate bond conductance in a pore network model. 
Comparing the SC (Figure \ref{fig:cubic}d) and the BCC arrays (Figure SF1 in the supplementary material) 
also demonstrates the significant effect of the packing of the media on the pore space, even for media with constant grain size.

\subsubsection*{Step 3b. $\lambda$-based medial axis partitioning--Finney medium} \label{sec:Finney}

\begin{figure} 
 \includegraphics[width=\textwidth]{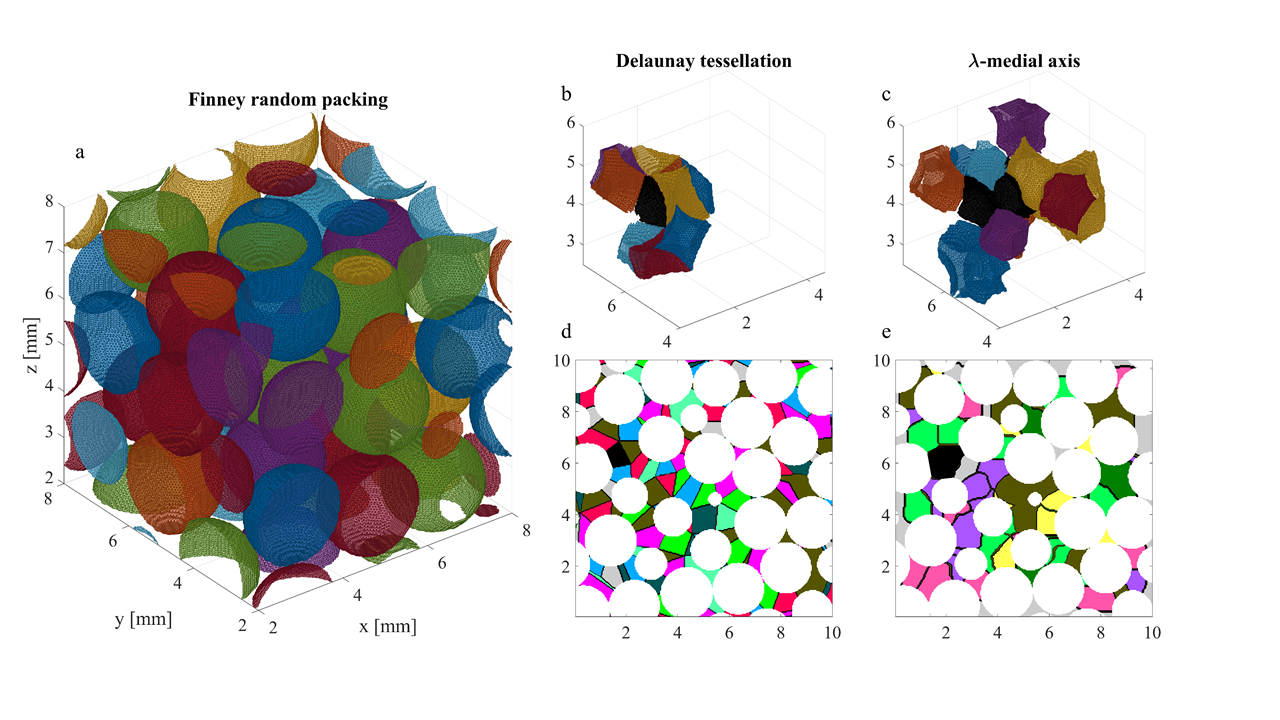}
 \caption{a) Finney spherical grains array. b) partitioned example pore and its neighbors using the Delaunay tessellation method. c) partitioned example pore and its neighbors using the $\lambda$-based medial axis method. d) a horizontal cross-section of the partitioned pore system using the Delaunay tessellation method. e) a horizontal cross-section of the partitioned (using the $\lambda$-based medial axis method) pore system. The cross-sections are depicted at an arbitrary height of $z=4.44 mm$ from the bottom. The $\lambda$ map was obtained using Gaussian filtering with $\sigma_0=2$, $\Delta \sigma=0.01$, and $\gamma=0.75$.} \label{fig:Finney}  
 \end{figure}

Partitioning of the pore space using the $\lambda$-based medial axis (Fig.~\ref{fig:Finney}c and e) is compared to the Delaunay tessellation and DM-watershed methods (Fig.~\ref{fig:Finney}b and d). The zoomed-in section depicted in Fig.~\ref{fig:Finney}b and c demonstrate how a large pore (in black) is "over-partitioned" from a pore-space morphology point of view by the Delaunay tessellation. Unlike morphology-centered methods (e.g., medial axis, watershed, and maximal balls), Delaunay tessellation is based on the grain locations and does not necessarily define the pore body size using its maximal inscribed sphere axis nor the throat (face) on its minimal one. As a result, the Delaunay tessellation underestimates this pore size and its connectivity compared to the $\lambda$-based medial axis partitioning. 
The pore size distributions using $\lambda$-based medial axis and Delaunay tessellation are presented in Figure SF3a and the pore volume distributions in Fig. SF3b in the supplementary material. The Delaunay tessellation over-partitions (from a morphologic point of view) large volume pores to a number of smaller ones. Moreover, as the pores are not circular, these large-volume pores can be separated (along their longer axis) into several large-diameter pores (defined in terms of the largest enclosed sphere). This explains the seemingly right shift in the peak of the pore size distribution (Fig. SF3a).    
This definition of a pore is inconsistent with the conceptualization of pore bodies and throats, allowing pore throats to be larger than the pore bodies. The effect of this inconsistency may be significant for the evaluation of a multiphase flow and transport and retention curves of the media \citep{alRaoush}. 
The importance of these differences between the partitioning methods depends on the flow problem at hand. For many flow problems, it is sufficient to evaluate the permeability; for others, one might be interested in the retention, or the flow velocities distribution.
For example, the large pore colored black in the cross-section shown in Fig.~\ref{fig:Finney}e is connected to ten pores (Fig.~\ref{fig:Finney}c, not all visible in the image point of view) and is more than double the size of the same pore when partitioned using the Delaunay tessellation method (Fig.~\ref{fig:Finney}e). Typically, in the Delaunay tessellation method, each pore has four throats (one on each side of the tetrahedron). However, in this example (Fig.~\ref{fig:Finney}e), the pore is connected to more than four pores, although some of these connections are through an extremely small (few pixels wide) interface. As a throat should technically not separate more than two pores, this means that the number of throats is not limited to the number of pore sides. The anomaly of a throat neighboring 3 or 4 pores is common for watershed algorithms and may suggest over-partitioning \citep{Prodanovic2006, Prodanovic2007}. On the other hand, for a complex 3D pore system, a throat-throat interface seems geometrically possible. 
For the Finney medium  (Fig.~\ref{fig:Finney}), the number of throats neighboring more than two pores (using 26-connectivity directions criteria) in the traditional DM-watershed method is twice as frequent as in the proposed $\lambda$-based medial-axis method. This may suggest that the $\lambda$-based medial-axis method resolves the watershed over-partitioning problem.  However, the Delaunay tessellation method has much fewer neighboring throats, probably because of its smaller and simpler-shaped pores (not shown). 

\subsection{Two-dimensional variably-saturated disordered granular media\label{sec:2d}}
Here, we use images from a two-dimensional millifluidic device that was used for multi-phase flow experiments \citep{jm2017}. In these experiments, the media are characterized by variable fluid saturation. The solid medium is composed of cylinders (pillars) with a mean diameter ($\langle d_g \rangle$) of $0.83$ mm (and standard deviation of $0.22$ mm), a height of $0.5$ mm, and a porosity of $0.71$. \citep{jm2017}. The pixel size of the 2D images is $\Delta x = 0.032$ mm.  The images do not include the entire original flow device and depict a flow domain of $L=105$ mm ($3292$ pixels) along the main flowing axis ($x$-axis) and width $W=70$ mm ($2182$ pixels).
We consider two images. The first, termed here {\em saturated}, corresponds to the saturated stage of the experiment before air was introduced to the system (Figure~\ref{fig:sat_2D_unfiltered}a). This medium is composed of disordered circles of varying diameters. The second, termed here {\em unsaturated}, corresponds to the partially-saturated case with an effective saturation degree $S_e=0.65$ (Figure~\ref{fig:2D_pb}c). $S_e$ is the porosity fraction filled with the continuous and percolated liquid phase. This medium comprises the union of the circular grains and the amorphous air phase (and also the disconnected water phase). 
PNM of the unsaturated case can be derived from the saturated pore space partitioning in union with the binary image of the phase configuration. However, this formulation disregards the wetting fluid film surrounding the grains and the actual connectivity, which may be significant for some consequences \citep{hoogland2016drainage}. 

\subsubsection*{Step 1. Distance and curvature maps}
Similarly to the regular spherical media described in the previous section, the DM of the saturated case can be evaluated exactly. However, the DM for the unsaturated case has to be evaluated numerically. Moreover, unlike the regular three-dimensional media, the disordered media (both saturated and unsaturated) may form plateaus in the DM, which in turn may lead to false pore identification. Thus, image analysis of these media types requires a filtering process, the level of which depends on the medium characteristics. This is discussed in detail in \ref{app:filtering}.

\subsubsection*{Step 2. Locating critical points}

\begin{figure} 
 \includegraphics[width=\textwidth]{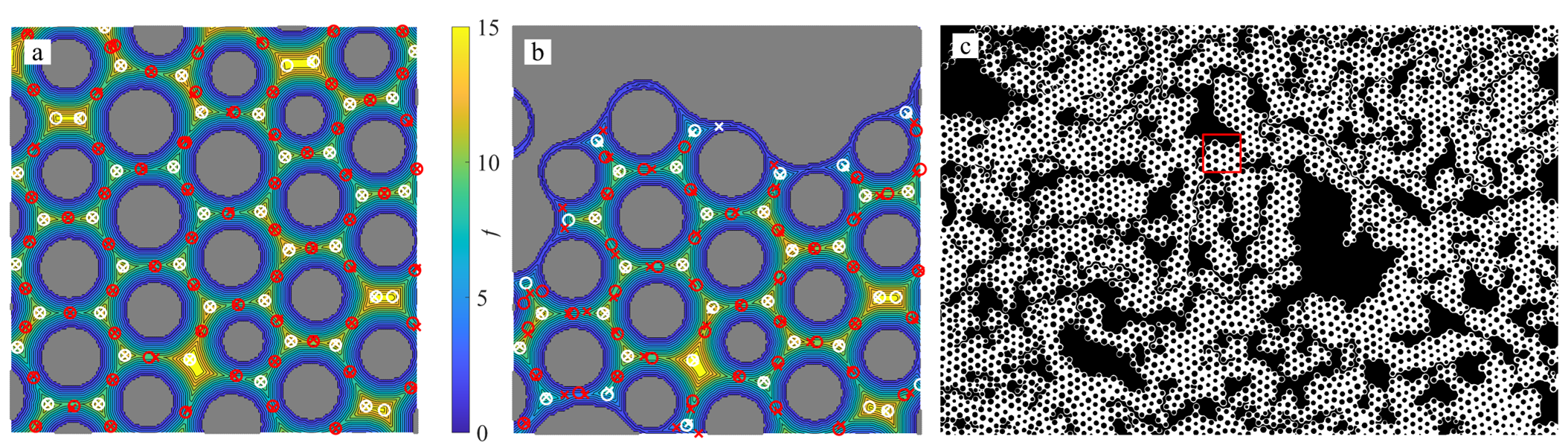}
 \caption{Distance map ($f$) and location of critical points using a Gaussian filter with $\sigma = 5$ [pixel] for a) the saturated and b) the unsaturated $2D$ cases. Pore throats' locations (best fit) are denoted by red stars, located saddle points are marked with red circles, pore bodies' locations (best fit) are denoted by blue stars, and the located peaks are marked with blue circles. Subplot c) presents the image of the unsaturated 2D media; the continuous percolated pore space is white, and the air, solid, and disconnected liquid are black. The best fit (stars) corresponds to the locations evaluated using a Gaussian pyramid and complementary dilution, using filtering parameters for the saturated and unsaturated cases of $\sigma _0=3$ and $0.5$ (in accordance), $\gamma =1$ and $0.75$ (in accordance), and $\Delta \sigma= 0.01$ (in both cases) \label{fig:2D_pb}}  
 \end{figure}

The critical points are located using a curvature map filtered once with a single layer Gaussian filter with $\sigma=5$, which corresponds to about $\sigma^2=\frac{\langle d_g\rangle}{\Delta x}$, and then with a Gaussian pyramid filter. The comparison is depicted in Figure \ref{fig:2D_pb}. For the saturated case (Fig.~\ref{fig:2D_pb}a), the single layer filtering works very well compared to the unfiltered case shown in Fig.~\ref{fig:sat_2D_unfiltered}a. However, for the unsaturated case, a single-layer Gaussian filter cannot correctly locate all the critical points (Fig.~\ref{fig:2D_pb}b). This is probably because of the much more complex and irregular pore system with widely varying pore sizes. In this case, which is more realistic for natural porous media, a pyramidal Gaussian filtering process is used to account for the different scales of the pore system (see ~\ref{app:filtering}).

\begin{figure}
 \includegraphics[width=\textwidth]{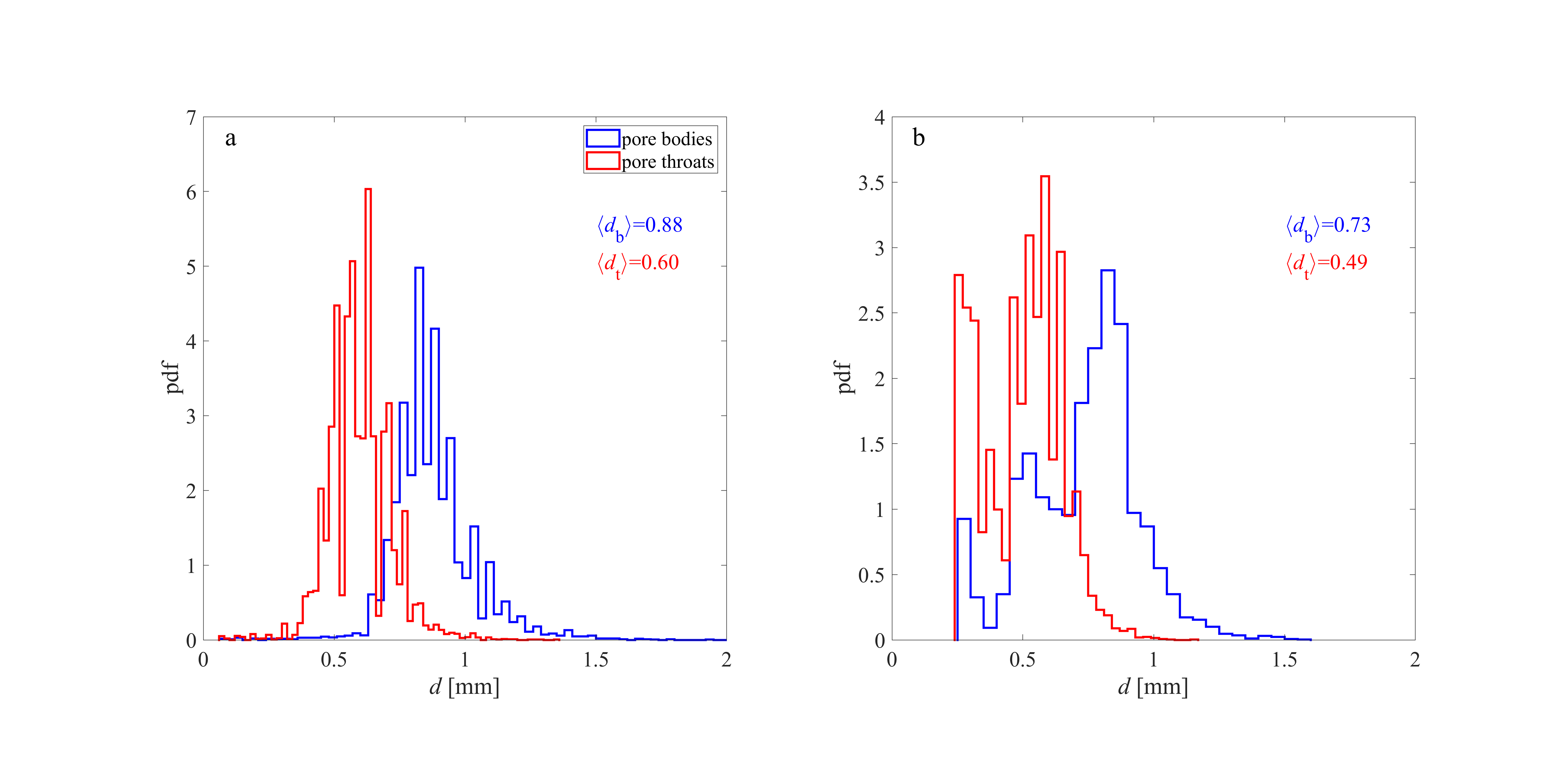}
 \caption{Pore bodies (blue) and throats (red) size distributions of the a) 2D saturated millifluidic device and b) 2D unsaturated ($S_e=0.65$) millifluidic device. The Gaussian filtering parameters used are $\sigma_0=2$, $\Delta \sigma=0.01$, and $\gamma=1$} \label{fig:psd}
 \end{figure}

Comparing the size distributions of pore throats and bodies of the unsaturated (Figure~\ref{fig:psd}b) to the saturated case (Figure~\ref{fig:psd}a) demonstrates how the invaded air affects the water-filled pore space. The mean pore size of the unsaturated case is smaller than for the saturated because air preferentially invades the larger pores due to capillary forces. Moreover, the pore size distribution of the unsaturated case has a ``dual'' distribution, represented by two local maximal points in the probability density function (Fig.~\ref{fig:psd}b). This can be explained by air invading large pores and splitting them into smaller ones, in which the residual water phase is reduced to a narrow film surrounding the grain. These narrow films have a very small radius and pose a great challenge to pore space characterization and locating methods.
It should be mentioned that if one disregards the very narrow water film pore size distribution, in which flow velocities are usually very small and its contribution to the permeability is negligible (except for very low saturation degrees), then a large $\sigma_0$ can be used to filter them out, or alternatively, eliminating criteria with a minimum pore size threshold. This example demonstrates the importance of the specific application and the target macroscale parameters when choosing the analysis method \cite{sheppard2006analysis}.

\subsubsection*{Step 3. $\lambda$-based watershed and medial axis partitioning} 

Here we demonstrate the $\lambda$-based medial axis (Figure~\ref{fig:watershed}a) and the $\lambda$-based watershed (Figure~\ref{fig:watershed}d). The results are compared to a traditional DM-based watershed method without complementary search algorithm (Figure~\ref{fig:watershed}b.) The partitioning process (regardless of the method) gives higher levels of information about the pore system compared to the locating of critical points, such as the pore areas and perimeters (or equivalently, the volume and surface area for 3D), which can be used to evaluate more detailed pore characteristics like the fractal dimension of the pore space \citep{blunt2017}.
Recently, the $\lambda$-based medial axis partitioning method was used to construct detailed pore network models that were found in excellent agreement with the flow estimation of direct numerical simulations, reflecting on the applicability of both network models and the image analysis~\citep{ben2024network}.
In order to compare the proposed partitioning methods and the traditional DM-watershed method, we determine the fractal dimension of the respective partitioned pore space and the sensitivity of the respective method to the image resolution.

\begin{figure}
 \includegraphics[width=\textwidth]{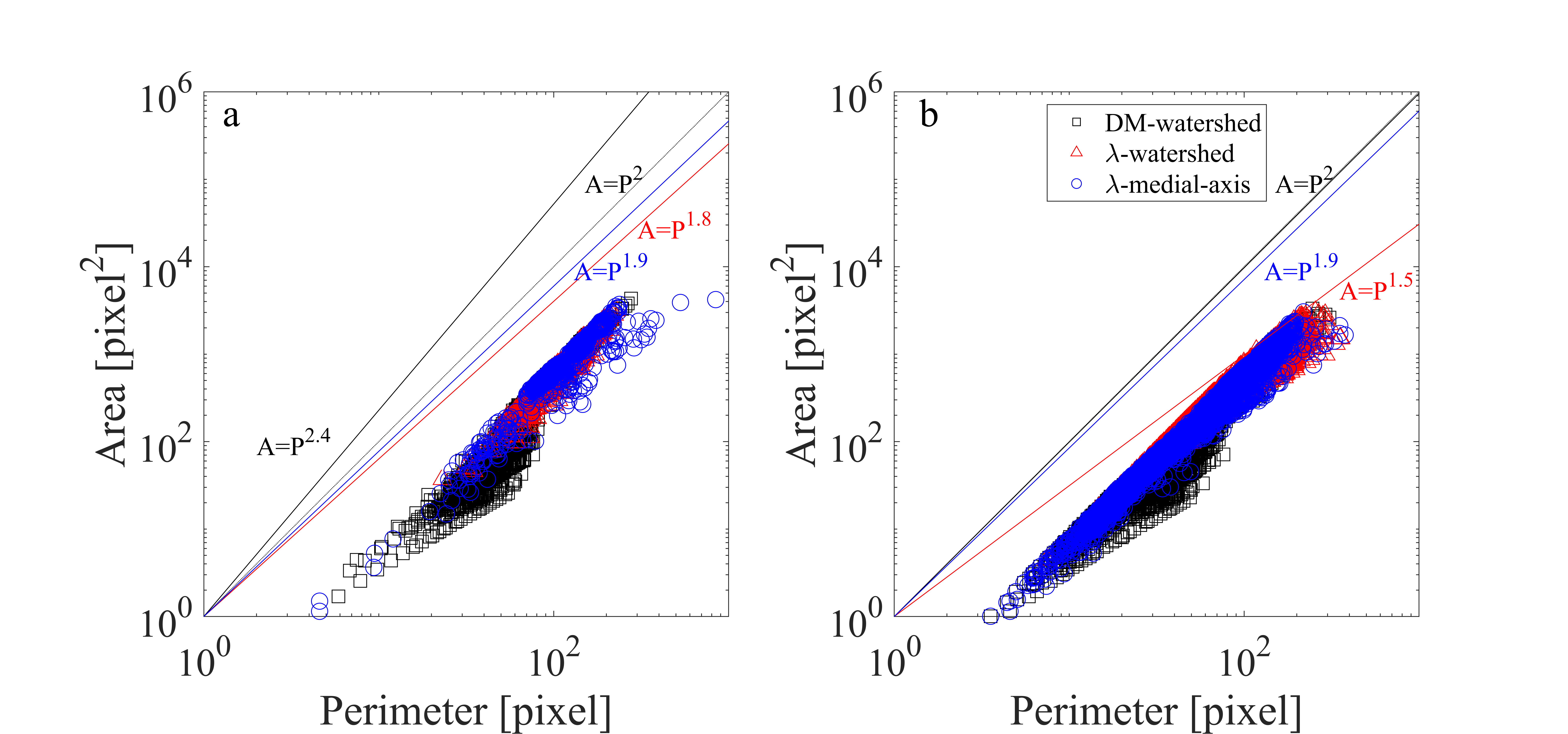}
 \caption{Fractal analysis of the pore space. Pore area as a function of the pore perimeter for a) saturated 2D granular media and b) unsaturated 2D porous media. partitioned pores were evaluated using the $\lambda$-based medial axis (blue circles, eq.~\ref{eq:lambda_1_binary}), $\lambda$-based watershed (red triangles), and the DM-based watershed (black squares). The slope of the Euclidean dimension is depicted with the broken black line, and the different partitioning methods' fractal dimension is depicted with the solid lines with the corresponding colors. $\lambda _1$ maps for the saturated and unsaturated cases where filtered using $\sigma _0=3$ and $0.5$ (in accordance), $\gamma =1$ and $0.75$ (in accordance), and $\Delta \sigma= 0.01$ (in both cases).} \label{fig:fractals} 
 \end{figure}

\paragraph*{Fractal dimension:} The fractal dimension $D$ of the pore space is defined here
through the dependence of the pore size (area $A$) and on the pore perimeter ($P$) as follows \citep{schlueter1997}
\begin{linenomath}
\begin{equation}\label{eq:fractal}
A \propto P^D.
\end{equation}
\end{linenomath}
%
The fractal analysis of the pore space of the saturated and unsaturated 2D cases is depicted in Figure \ref{fig:fractals}. For the saturated case (Fig.~\ref{fig:fractals}a), both newly-presented $\lambda$-based methods give a fractal dimension close (with $D<2$) to the Euclidean dimension (broken black line, Fig.~\ref{fig:fractals}a) denoted by the logarithmic relation between the pore perimeter and its area. This suggests that the size of the pores does not affect their shape, which is to be expected in a granular media. The $\lambda$-based medial axis (blue circles) tends to unite pores with large throats. Thus, it defines slightly larger pores. This analysis illustrates the predominance of these methods over the DM-based watershed (black squares, Fig.~\ref{fig:fractals}), which results in a higher fractal dimension and a gross over-partitioning, visible in the number and shape of the small false pores in the saturated (Fig.~\ref{fig:fractals}a) case. For the unsaturated case (Fig.~\ref{fig:fractals}b), the medial axis (blue circles) partitions the narrow water films surrounding the grains into extremely small pores, which interestingly maintain the same fractal dimensionality (Fig.~\ref{fig:fractals}b). These small pores are excluded from the $\lambda$-based watershed (red triangles) when it is processed using the critical points locating method (annexation of pores without a pore body), using the prescribed filtering parameters ($\sigma _0=0.5$, $\gamma =0.75$, and $\Delta \sigma= 0.01$). In this context, the $\lambda$-based medial axis is less subjective to filtering parameters and might be superior for complex media. This is illustrated in the much different fractal dimension in the unsaturated case. For the unsaturated case, the DM-based watershed partitioning gives a fractal dimension close to the Euclidean dimension and much smaller than the value obtained for the saturated case. The change in the fractal dimension may suggest that this method is also very sensitive to the small and narrow water film pores. 

\begin{figure}
 \includegraphics[width=\textwidth]{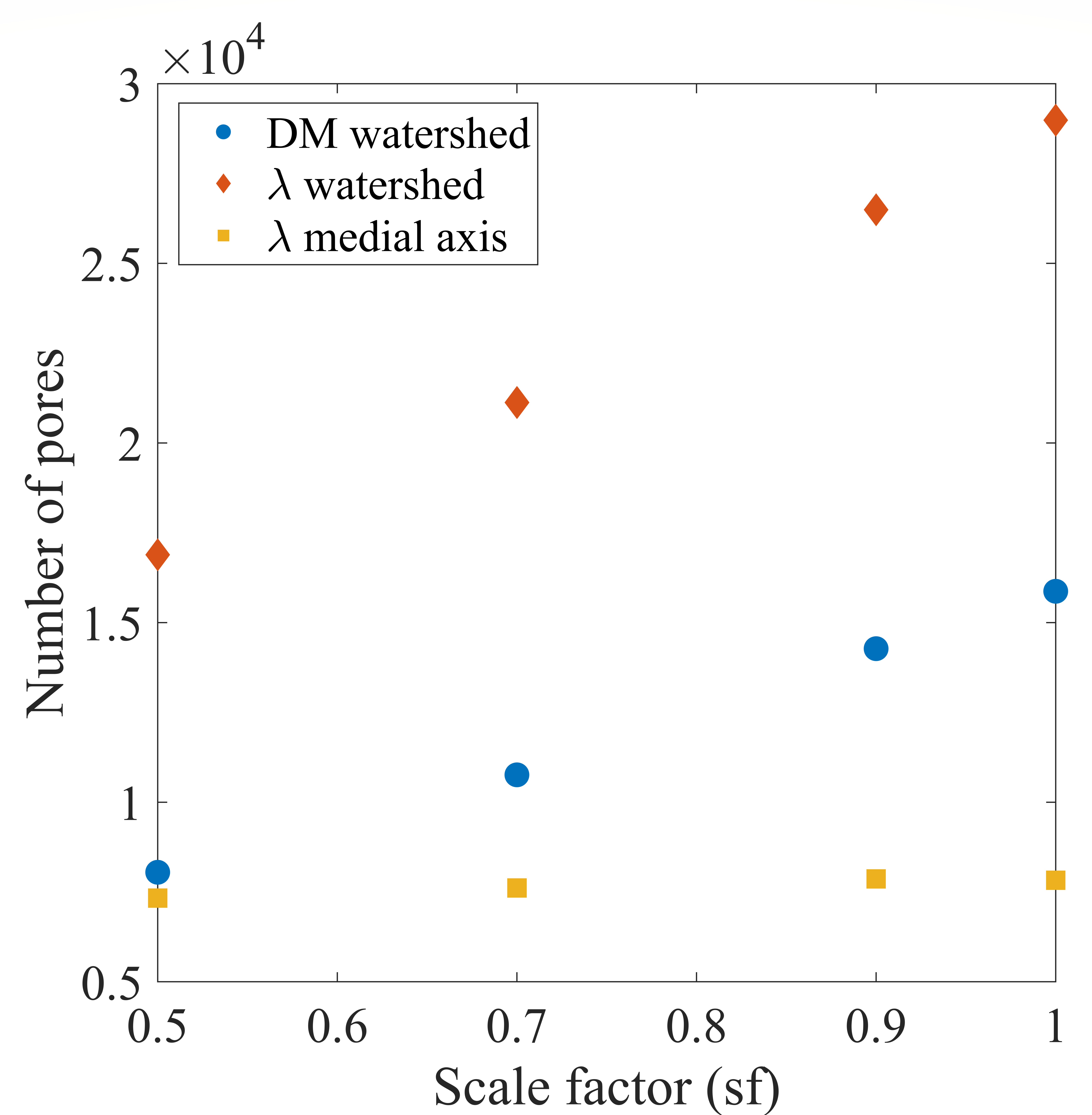}
 \caption{Effect of the image resolution, determined by the scaling down factor (sf) on the number of partitioned pores using the different partitioning method for the unsaturated 2D media.} \label{fig:Resolution2D} 
 \end{figure}

\paragraph*{Sensitivity to image resolution:} The sensitivity of the different methods to the image resolution is evaluated by running the partitioning algorithm on different levels of scaled-down binary images. The scaling down factor, denoted by $sf$, represents the ratio between the sides of the original pixels to the scaled-down pixels, for example for $sf=0.5$ the pixels of the scaled-down images are double the size of the original pixels. The values in the pixels of the scaled binary image (i.e., 0 or 1) are set using the rounding of a bicubic interpolation. The $\lambda$-based medial axis is significantly less sensitive to the image resolution in terms of the number of pores compared to the watershed methods Figure~\ref{fig:Resolution2D}. This is mostly because of the over-partitioning of small pores at the thin water films ($\lambda$-based watershed) and at the false saddles (DM-based watershed). 

\subsection{Three-dimensional complex rock samples}\label{sec:berea}

Here, we apply the $\lambda$-based medial axis method for pore-space partitioning in three different rock samples, a Berea and a Bentheimer sandstone sample \citep{berea} as well as a Tight Carbonate sample \citep{Carbonate}. In the following, we exemplify the method for the Berea sandstone and compare its performance to the DM-watershed method. Binary images of a Berea sandstone (Figure~\ref{fig:Berea_body_throat}a) are obtained from \cite{berea}. The images present a cubic rock sample with a volume of $~0.125$mm$^3$ with a resolution of $2.25 \mu$m (i.e., $223$ pixels on each side). The binary images and images of the partitioned pore space for the Bentheimer sandstone and Tight Carbonate samples can be found in the supplementary material. 

\subsubsection*{Step 1. Distance and curvature maps}
The Berea sandstone sample's DM must be evaluated numerically from the binary image because we do not have any prior information about the grains' or pores' shape geometry. The DM and determinant of the Hessian matrix maps are depicted in Figure SF2 in the supplementary material. The Berea sandstone has a complex pore system (Figure~\ref{fig:Berea_body_throat}a) and a low porosity of $0.19$ \citep{berea}, which makes it a challenging benchmark for pore system characterization methods. Processing this complex system requires a Gaussian pyramid filtering as described in \ref{app:filtering}.

\subsubsection*{Step 2. Locating critical points} 

\begin{figure}
   \includegraphics[width=\textwidth]{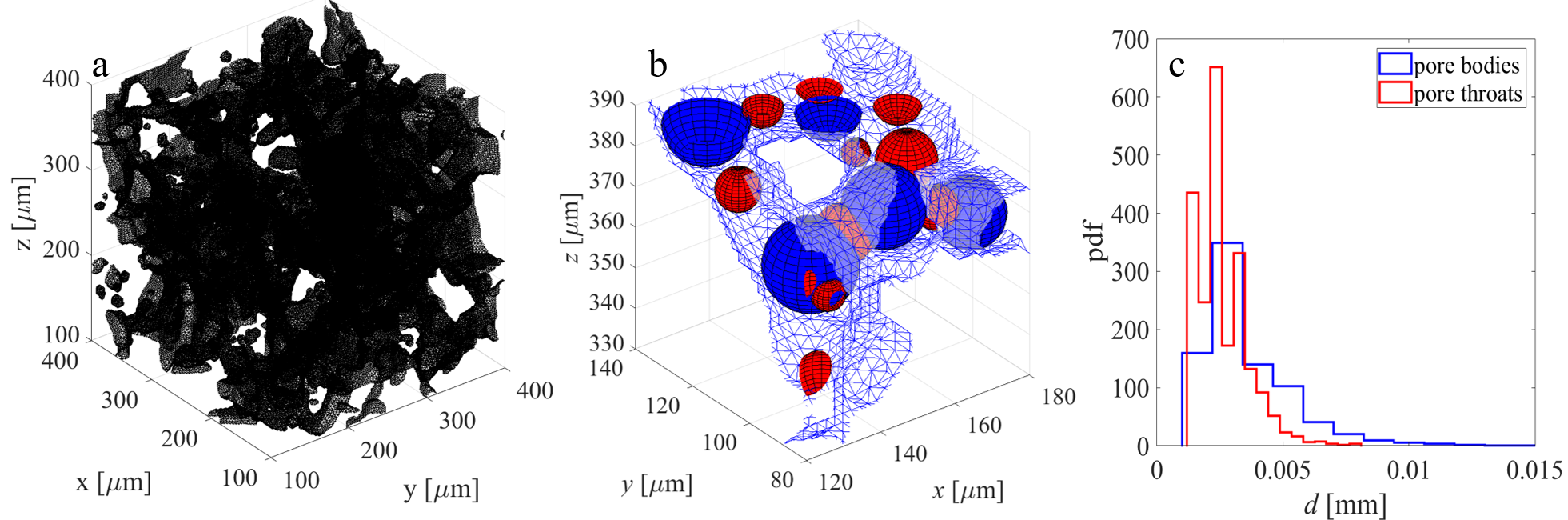}
   \caption{a) Binary image of a Berea sandstone (taken from \protect\cite{berea}). b) A zoomed-in image of the Berea sandstone pore space. Blue spheres represent the pore body center and diameter, and red spheres represent the pore throats' locations and diameters. c) Pore bodies (blue) and throats (red) size distributions of the Berea sandstone. Image was processed with filtering parameters of $\gamma=0.75$, $\sigma_{0}=2$, $\Delta \sigma=0.01$.} \label{fig:Berea_body_throat} 
 \end{figure}

Figure~\ref{fig:Berea_body_throat}b presents a zoomed-in section of the rock and the locations of the body (blue) and throat (red) centers. The complexity of the pore system and the required image resolution of real rock samples is illustrated in the depicted section, with a volume of $60^3$ $\mu$m$^3$ that includes five pores centers and about ten throats. 
Unlike in the two-dimensional variably-saturated disordered granular media and the regular three-dimensional media, in the Berea sandstone, the pore bodies and throats have very similar sizes (Figure~\ref{fig:Berea_body_throat}c). This may explain, for example, the relatively weak hysteretic behavior of Berea sandstone \citep{pentland2011}. The mean coordination number of the Berea sandstone, based on twice the number of throats to pores ratio, found by the locating method is $\langle Z \rangle = 2.97$, which is very close to the value reported by \cite{yanuka1986} of $\langle Z \rangle=2.9$.  However, this coordination number accounts for many connected in-series ($Z=2$) and dead-end ($Z=1$) pores along the fractures (chain of connected in-series pores). While these dead-end and in-series connected pores may affect transport phenomena and retention, they obscure the interpretation of the coordination number as a parameter of the flowing phase connectivity. In Berea sandstone, to account for the relevant-to-the-flow pore connectivity, the mean coordination number can be defined as the number of edges leading into each vertex, that is, accounting only for pores with a larger coordination number than two \citep{lin1982}. This analysis, however, requires partitioning.  

\subsubsection*{Step 3. $\lambda$-based medial axis partitioning}

\begin{figure}
 \includegraphics[width=\textwidth]{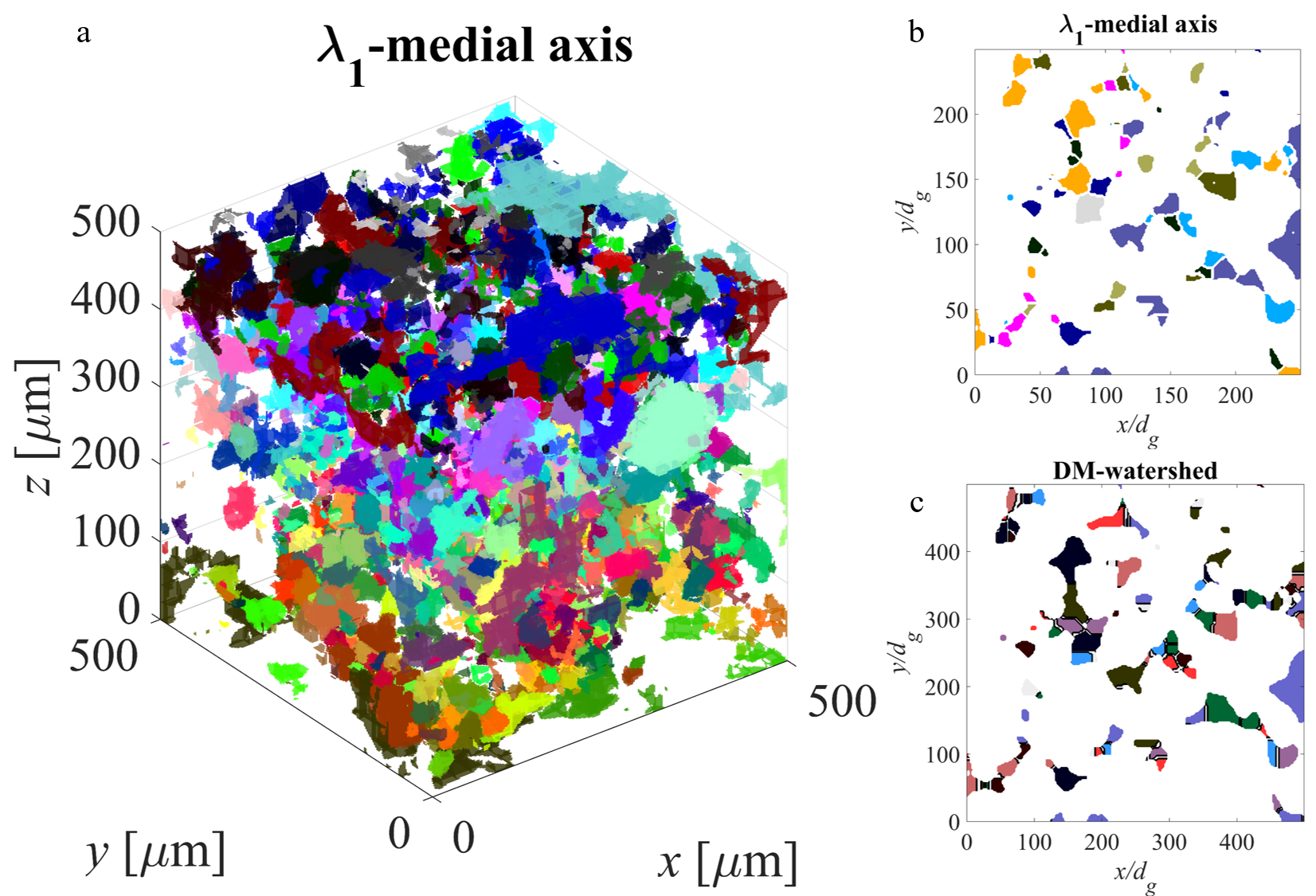}
 \caption{a) partitioned binary image of a Berea sandstone using the $\lambda$-based medial axis, b) cross-section of the partitioned pore space using the $\lambda$-based medial axis, and c) cross-section of the partitioned pore space using the distance map-based watershed. the cross-sections are taken at the mid-height of the Berea sandstone sample. $\lambda_1$ map was obtained by processing the binary image with filtering parameters of $\gamma=0.75$, $\sigma_{0}=2$, $\Delta \sigma=0.01$.} \label{fig:Berea_seg}  
 \end{figure}

Figure \ref{fig:Berea_seg}a presents the partitioning of the Berea sandstone using the $\lambda$-based medial axis method described in Section \ref{sec:lma}. This method was chosen here over the $\lambda$-based watershed method because it is more objective regarding the filtering processes. As discussed at the end of Section~\ref{sec:2d}, the medial axis relies only on the $\lambda$-skeleton, which, unlike the watershed method, is not sensitive to the precise location of the critical points. As in the previous section, we compare the $\lambda$-based medial axis to the traditional DM-watershed method. 

It is visible that the pores of the Berea sample have a very irregular and highly angular shape.
When comparing the cross-sections (at mid-height) of the partitioned pore system of the Berea sandstone using the $\lambda$-based medial axis (Figure \ref{fig:Berea_seg}b) and DM-watershed (Figure \ref{fig:Berea_seg}c), it seems that the latter is over-partitioned at some throats. This over-partitioning manifests itself in the narrow and elongated shape of some of the small pores wedged between larger pores, which are not found in the $\lambda$-based medial axis in the sense that not all these locations separate between pores.  

In analogy to the 2D fractal analysis (Eq.\ref{eq:fractal}), the fractal dimension is defined through the relation between pore volume $V$ and the surface area $S$ of the pore,
\begin{linenomath}
\begin{equation}\label{eq:fractal3D}
V \propto S^{\frac{D}{2}},
\end{equation}
\end{linenomath}

\begin{figure}
 \includegraphics[width=\textwidth]{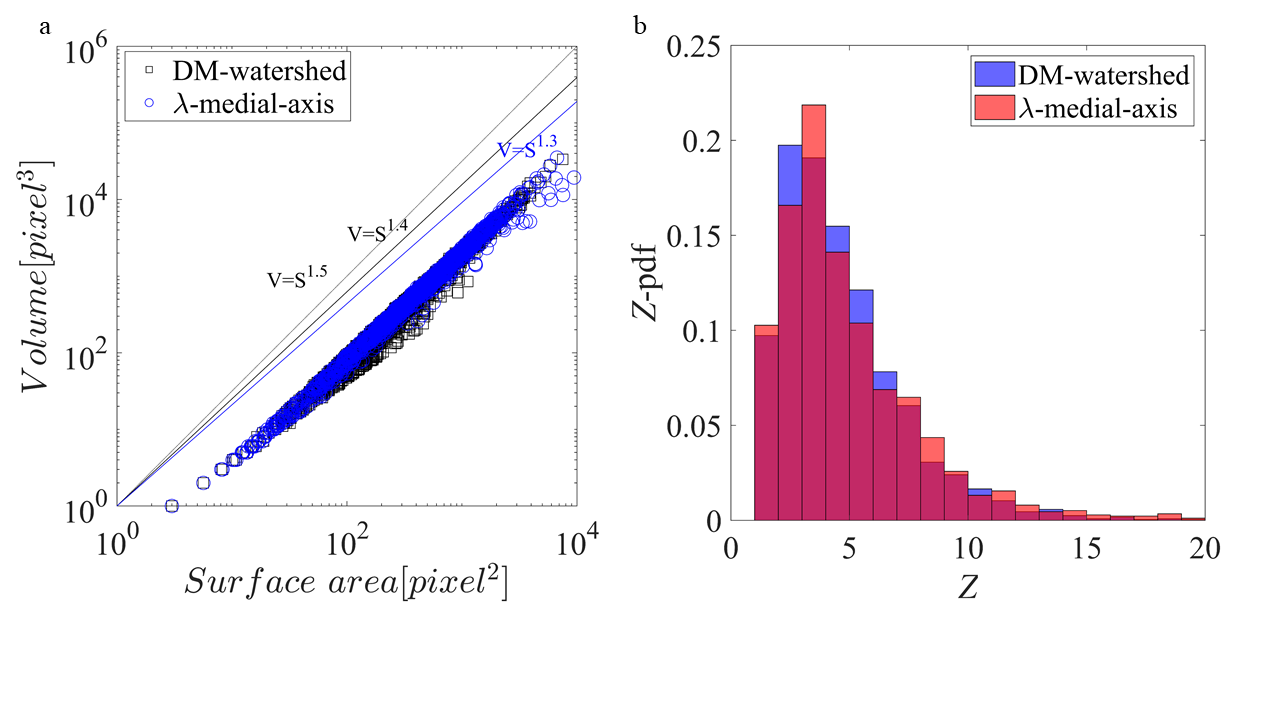}
 \caption{a) Fractal analysis of the pore bodies. The pore volume ($V$) as a function of the pore surface area ($S$) and b) the coordination number probability density function of the partitioned pore space using different partitioning methods. The broken line in (a) represents the Euclidean fractal dimension (3) slope, and the solid lines the fractal dimensions evaluated by the different partitioning methods} \label{fig:Berea_fractalAndZ}  
 \end{figure}

 Despite the irregular and angular shape of the pores, the fractal dimension of the pores is similar to the Euclidean dimension (broken line in Figure \ref{fig:Berea_fractalAndZ}a), regardless of the partitioning method. The span of pore volumes and areas of the different partitioning methods is similar (except for very few large pores in the $\lambda$-based medial axis, probably due to some boundary effect on the $\lambda_1$ map). However, the DM-based watershed partitioning seems to have a wider shape span at the mid-size pores, with a higher surface area to volume ratio (i.e., more ``spherical'').  The coordination number distribution of the partitioned pore system is very similar in both methods ($\lambda$-based medial axis vs. DM-based watershed). However, there are noticeably more $Z=2$ in the DM-watershed than in the $\lambda$-based medial axis (Figure~\ref{fig:Berea_fractalAndZ}b) because of the over-partitioning of the marker-based watershed methods. The mean coordination number of the partitioned pore system ($\approx 4.5$) is much larger than the value evaluated by twice the throat-to-bodies ratio. This is because in a complex 3D medium, a throat often separates more than two pore bodies (Figure \ref{fig:Berea_seg}b), see discussion in section \ref{sec:Finney}. This explanation is supported by the similar number of pores found by the critical point locating method ($1,875$) and the $\lambda$-based medial axis partitioning ($1,818$).   

Images of the partitioned pore space for the Bentheimer sandstone and Tight Carbonate sample are depicted in Fig. SF4 in the supplementary material. In both cases, we used the same filtering parameters despite the significant differences between the image resolution and pore sizes. The Bentheimer Sandstone sample was obtained at a resolution of 2.25 $\mu$m and analyzed in a volume of $223 \times 223 \times 223$ pixels, while the Tight Carbonate was obtained at a resolution of 30.6 $\mu$m voxel's length and a volume of $500 \times 500 \times 500$ pixels. This point demonstrates the robustness of the method.  

\section{Summary and conclusion}

Modeling processes in granular and porous systems hinges on accurately describing the pore system. This holds for many applications, spanning from the interpretation of phase distribution obtained from flow cells, microfluidic devices, and rock samples (e.g., via X-ray microtomography images) to developing pore network models to evaluate flow and transport through porous media. We presented a binary image processing framework for pinpointing pore bodies and throat centers and characterizing granular and porous media from binary images. The new approach uses the curvature map of the distance transform rather than the distance map for the location and partitioning process. The proposed method is robust and can be used to directly construct simple pore network models. We use this framework to derive two new partitioning methods based on the watershed and medial axis approaches. 

The $\lambda$-based watershed method uses the map of the minimum eigenvalue of the Hessian matrix rather than the distance map as the intensity image map. The suggested method significantly mitigates the common problem of saddle-induced over-partitioning shared by all traditional marker-based watershed methods. The $\lambda$-based medial axis method again uses the map of the minimum eigenvalue. This method first creates a binary image and then uses the skeletonization of this image in union with the original void space binary image to easily differentiate between distinct pores. Validation and demonstration of these methods for media of different complexity show promising results in both accuracy and robustness. Future research can use mercury-porosimetry to compare the capabilities of morphology-centered methods, such as the ones suggested in this work, to other methods, such as Delaunay tessellation for evaluating different media retention.

The main limitation of the proposed approach is the excessive filtering process required for the more complex media. On the other hand, this filtering process allows some agility in interpreting the relevant topology. For example, if one does not care about the smallest pores to evaluate the media permeability, one can choose filtering parameters to eliminate these features. In this context, a future research path can relate the filtering parameters to the medium properties and required target parameters and processes.


 \section*{Acknowledgments}
The authors thank Joaquin Jimenez-Martinez for the images of the granular media, and to three anonymous reviewers and Professor Maša Prodanovic for highly constructive comments. 
I.B.N and M.D acknowledge funding from the European Union’s Horizon 2020 research and innovation program under the Marie Sklodowska-Curie grant agreement No. 101066596 (USFT). 
J.J.H. acknowledges the support of the Spanish Research Agency (10.13039/501100011033), Spanish Ministry of Science, Innovation, and Universities through grant HydroPore II PID2022-137652NB-C41.
M.D. acknowledges funding from the European Union (ERC, KARST, 101071836).

 \section{Declarations}
 \paragraph*{Conflict of interest} 
 The authors have no relevant financial or non-financial interests to disclose.
 
 \paragraph*{Data Availability}
A MATLAB function of this algorithm and a sample binary image are provided in a data repository with a permanent identifier \cite{ben_noah_2024_10633605}. This function requires a MATLAB imaging toolbox extension and was tested on the 2022b version.


\begin{appendices}

\section{Gaussian filters}\label{app:filtering}
Gaussian filtering is commonly used for image analysis.  The major merit of the Gaussian filter is that it does not create new artificial extremum points. 

\emph{Single-layer Gaussian filtering.}

The Gaussian filter smooths the image by convolving the input signal (distance map) with a Gaussian function, which for 2D is written as
\begin{linenomath}
\begin{equation}\label{eq:Gauss_filter}
G(x,y)=\frac {1}{2\pi \sigma ^{2}}e^{-(x_1^{2}+x_2^{2})/(2\sigma ^{2})},
\end{equation}
\end{linenomath}
where $\sigma$ is the standard deviation of the Gaussian, and $x_1$ and $x_2$ are the distances from the original pixel in perpendicular directions. 
In the convolution, the values of the output filtered distance map at each point are affected most strongly by the original value at this point and by the values on near pixels with a weight that is declining with the distance (according to the Gaussian, \eqref{eq:Gauss_filter}.
In this sense, applying a Gaussian filter is similar to running the heat (or diffusion) equation on the distance map, as it cannot introduce new extreme points. This feature is of major importance for the application of a critical point localization algorithm. Here we use the Gaussian filtering functions implemented in the MATLAB imaging toolbox~\citep{MATLAB}. 
  
\emph{Gaussian pyramid filtering}

The single-layer Gaussian filtering can effectively reduce the number of false points for regular media such as the saturated granular 2D media (Fig.\ref{fig:sat_2D_unfiltered}a). However, for more realistic porous media, more potent filtering might be needed. Gaussian pyramid is a method where an image is subjected to repeated smoothing and sub-sampling, where the smoothed image is iteratively stacked atop the previous (lower) level image.  The subsequent images are scaled and weighted down using a Gaussian average so that each pixel contains a local average corresponding to a neighborhood pixel from a previous iteration level image (lower level of the pyramid). During the filtration process, at each pyramid level, the Hessian matrix, its determinant and eigenvalues $\lambda$ are calculated from the filtered distance map, using different standard deviation values ($\sigma$), and weighted to form a stacked determinant and $\lambda$ maps. Here we use a constant $\sigma$ intervals ($\Delta \sigma$) so that at the $k$-th iteration (pyramid level)
\begin{linenomath}
\begin{equation}\label{eq:sigma}
\sigma_{k}=\sigma_0+(k-1)\Delta \sigma, \;\; k=1,2,...N_k,
\end{equation} 
\end{linenomath}
where $N_k$ is the number of iterations or pyramid levels, and $\sigma_0$ is the minimal (pyramid base level) standard deviation value for the Gaussian filtering.  

Then the summed filtered Hessian determinant ($D$) is calculated so that different pyramid levels are weighted as followed 
\begin{linenomath}\begin{equation}\label{eq:scale_space}
D_{k}=D_{k-1}+\frac{1}{2}{\sigma}_{k}^{\gamma/2} \cdot det(H_{k}), \;\; k=1,2,...N_{k}
\end{equation}
\end{linenomath}
where $\gamma \in [0,1]$ is the scale-space weighting factor of the pyramid, related to the dimensionality of the image feature \citep{lindeberg1999}, and $H_{k}$ is the Hessian matrix of the distance map filtered with $\sigma_{k}$. For large $\sigma_k$, the filtered distance map is getting extremely blurred to no longer affect the number or location of the critical points. 
This means that a stopping criteria ($N_k$) can be obtained from the filtered image or from a large enough maximal $\sigma$ value. 
The number of needed pyramid layers depends on the media, $\Delta \sigma$, and the desired resolution.

\emph{Error estimates and sensitivity analysis}

We investigate two aspects of the accuracy of the proposed methods: i) finding the correct number of critical points and ii) finding the exact spatial location of these points. To quantify these aspects, we define two magnitudes: the relative number of false critical points and the spatial error.
The relative number of false (or missing) critical points $E_{N,i}$ (where $i=s$ for saddle and $i=p$ for peaks) is defined as the relative difference between the number of critical points located by the algorithm $N_i$ and the actual (or best fitted, i.e., after the complementary dilution and compared to the number of pores evaluated from the partitioning methods) number $N_{j}$. That is
\begin{linenomath}
\begin{equation}\label{eq:Number_error}
  E_{N,i} = \frac{N_{i} - N_{j}}{N_{j}},
\end{equation}
\end{linenomath}
where  $i = s; j=t$ corresponds to saddle points and pore throat, and $i= p; j=b$ corresponds to peaks and pore bodies, as appropriate. Notice that $E_N$ attains negative values when the locating method underestimates the number of throats and positive values for overestimation.
The spatial error $E_i$ is defined as the ratio between non-accurate to accurate critical points, where an accurate point is defined as a point located at a distance from a real critical point that do not exceed $\Delta x$, that is,
\begin{linenomath}
\begin{equation}\label{eq:Spatial_error}
E_i=\frac{\sum_{j=1}^{N_j}{\mathbb I\left(\min\left(\lVert \mathbf{x}_j-\mathbf{x}_i \rVert\right) > \Delta x\right)}}{\sum_{j=1}^{N_j}{\mathbb I\left(\min\left(\lVert \mathbf{x}_j-\mathbf{x}_i \rVert\right) \leq \Delta x\right)}},
\end{equation}
\end{linenomath}
where $\mathbf{x}_j$ real throats ($j=t$) or bodies ($j=b$), $N_j$ the number of throats or bodies, and $\mathbf{x}_i$ are the coordinates of the located saddle ($i=s$) or peak ($i=p$) points.

The proposed filtering method uses three independent parameters (at the maximum), the base level Gaussian standard deviation ($\sigma_{0}$), the standard deviation increment between adjacent levels of the Gaussian pyramid ($\Delta \sigma$), and the space scale factor ($\gamma$). At the same time, the noise and relevant scale of the filtering process are related to the image resolution and media properties (e.g., pore size distribution). Figure (SF5)
presents the parameters' sensitivity analysis of the 2D media. For the irregular granular media case, $E_{N,s}$ is close or equal zero for $1<\sigma_{0}$ and almost unaffected by $\Delta \sigma$ or $\gamma$ (Fig.SF5a). The insignificant effect of $\gamma$ is somewhat surprising as, in general, the detection scale using $\gamma=1$ is twice as large as for $\gamma=0.75$, which usually results in more shape distortions and a lower ability to capture throats in disordered media \citep{lindeberg1999}. Furthermore, in all the analyzed scenarios, $E_{N,s}$ is non-negative, i.e., the suggested method is either accurate or overestimates the number of throats. This merit reflects on the usability of the complementary saddle points dilution process, as described in the subsection \ref{sec:Complementary}. 

The spatial inaccuracy ($E_i$) is affected by the image resolution, where the accuracy decreases with the reduction in resolution (not presented). 
Moreover, in the saturated case, the effect of $\sigma_{0}$ is non-monotonic such that both too small or too large $\sigma_{0}$ may cause an error in the spatial accuracy of locating saddle points (Fig. SF5b). Despite that, this method shows robustness in the sense that a wide range of $\sigma_{0}$ was found to have a small spatial error (Fig. SF5b). 

For the unsaturated $2D$ media, as in the saturated case, $\gamma=1$ seems to be slightly superior to $\gamma=0$ and similar to $\gamma=0.75$.  The effect of $\sigma_0$ on the number of false points (Fig. SF5c) is very similar to the saturated case (Fig. SF5a) with any $1<\sigma_0<5$ seems to be adequate. On the other hand, increasing $\sigma_0$ monotonically decreases the spatial accuracy of the saddle points $E_s$ (Fig. SF5d). Interestingly, $\Delta\sigma$ slightly increases the number of false points ($E_{N,s}$) and, at the same time, slightly reduces the spatial error ($E_{s}$). 

Figure SF6 (in the supplementary material presents the effect of the filtering parameters on the pore throat radius distribution for the Finney random packing media. In general, the fraction of small throats decreases with increasing $\sigma_0$. The throat distribution is not sensitive to $\gamma$ and $\Delta\sigma$, in the analyzed range, with the exception of the $\sigma_0=1$, $\Delta\sigma=0.001$ and $\gamma=1$ case (red line in the top left panel).    

\end{appendices}


\bibliography{sn-bibliography}


\begin{thebibliography}{58}
\ifx \bisbn   \undefined \def \bisbn  #1{ISBN #1}\fi
\ifx \binits  \undefined \def \binits#1{#1}\fi
\ifx \bauthor  \undefined \def \bauthor#1{#1}\fi
\ifx \batitle  \undefined \def \batitle#1{#1}\fi
\ifx \bjtitle  \undefined \def \bjtitle#1{#1}\fi
\ifx \bvolume  \undefined \def \bvolume#1{\textbf{#1}}\fi
\ifx \byear  \undefined \def \byear#1{#1}\fi
\ifx \bissue  \undefined \def \bissue#1{#1}\fi
\ifx \bfpage  \undefined \def \bfpage#1{#1}\fi
\ifx \blpage  \undefined \def \blpage #1{#1}\fi
\ifx \burl  \undefined \def \burl#1{\textsf{#1}}\fi
\ifx \doiurl  \undefined \def \doiurl#1{\url{https://doi.org/#1}}\fi
\ifx \betal  \undefined \def \betal{\textit{et al.}}\fi
\ifx \binstitute  \undefined \def \binstitute#1{#1}\fi
\ifx \binstitutionaled  \undefined \def \binstitutionaled#1{#1}\fi
\ifx \bctitle  \undefined \def \bctitle#1{#1}\fi
\ifx \beditor  \undefined \def \beditor#1{#1}\fi
\ifx \bpublisher  \undefined \def \bpublisher#1{#1}\fi
\ifx \bbtitle  \undefined \def \bbtitle#1{#1}\fi
\ifx \bedition  \undefined \def \bedition#1{#1}\fi
\ifx \bseriesno  \undefined \def \bseriesno#1{#1}\fi
\ifx \blocation  \undefined \def \blocation#1{#1}\fi
\ifx \bsertitle  \undefined \def \bsertitle#1{#1}\fi
\ifx \bsnm \undefined \def \bsnm#1{#1}\fi
\ifx \bsuffix \undefined \def \bsuffix#1{#1}\fi
\ifx \bparticle \undefined \def \bparticle#1{#1}\fi
\ifx \barticle \undefined \def \barticle#1{#1}\fi
\bibcommenthead
\ifx \bconfdate \undefined \def \bconfdate #1{#1}\fi
\ifx \botherref \undefined \def \botherref #1{#1}\fi
\ifx \url \undefined \def \url#1{\textsf{#1}}\fi
\ifx \bchapter \undefined \def \bchapter#1{#1}\fi
\ifx \bbook \undefined \def \bbook#1{#1}\fi
\ifx \bcomment \undefined \def \bcomment#1{#1}\fi
\ifx \oauthor \undefined \def \oauthor#1{#1}\fi
\ifx \citeauthoryear \undefined \def \citeauthoryear#1{#1}\fi
\ifx \endbibitem  \undefined \def \endbibitem {}\fi
\ifx \bconflocation  \undefined \def \bconflocation#1{#1}\fi
\ifx \arxivurl  \undefined \def \arxivurl#1{\textsf{#1}}\fi
\csname PreBibitemsHook\endcsname

\bibitem[\protect\citeauthoryear{Arag{\'o}n-Calvo et~al.}{2007}]{aragon2007}
\begin{barticle}
\bauthor{\bsnm{Arag{\'o}n-Calvo}, \binits{M.A.}},
\bauthor{\bsnm{Jones}, \binits{B.J.}},
\bauthor{\bsnm{Van De~Weygaert}, \binits{R.}},
\bauthor{\bsnm{Van Der~Hulst}, \binits{J.}}:
\batitle{The multiscale morphology filter: identifying and extracting spatial
  patterns in the galaxy distribution}.
\bjtitle{Astronomy \& Astrophysics}
\bvolume{474}(\bissue{1}),
\bfpage{315}--\blpage{338}
(\byear{2007})
\end{barticle}
\endbibitem

\bibitem[\protect\citeauthoryear{Armstrong et~al.}{2019}]{armstrong}
\begin{barticle}
\bauthor{\bsnm{Armstrong}, \binits{R.T.}},
\bauthor{\bsnm{McClure}, \binits{J.E.}},
\bauthor{\bsnm{Robins}, \binits{V.}},
\bauthor{\bsnm{Liu}, \binits{Z.}},
\bauthor{\bsnm{Arns}, \binits{C.H.}},
\bauthor{\bsnm{Schl{\"u}ter}, \binits{S.}},
\bauthor{\bsnm{Berg}, \binits{S.}}:
\batitle{Porous media characterization using minkowski functionals: Theories,
  applications and future directions}.
\bjtitle{Transport in Porous Media}
\bvolume{130},
\bfpage{305}--\blpage{335}
(\byear{2019})
\end{barticle}
\endbibitem

\bibitem[\protect\citeauthoryear{Al-Raoush et~al.}{2003}]{alRaoush}
\begin{barticle}
\bauthor{\bsnm{Al-Raoush}, \binits{R.}},
\bauthor{\bsnm{Thompson}, \binits{K.}},
\bauthor{\bsnm{Willson}, \binits{C.S.}}:
\batitle{Comparison of network generation techniques for unconsolidated porous
  media}.
\bjtitle{Soil Science Society of America Journal}
\bvolume{67}(\bissue{6}),
\bfpage{1687}--\blpage{1700}
(\byear{2003})
\end{barticle}
\endbibitem

\bibitem[\protect\citeauthoryear{Bultreys et~al.}{2016}]{bultreys2016imaging}
\begin{barticle}
\bauthor{\bsnm{Bultreys}, \binits{T.}},
\bauthor{\bsnm{De~Boever}, \binits{W.}},
\bauthor{\bsnm{Cnudde}, \binits{V.}}:
\batitle{Imaging and image-based fluid transport modeling at the pore scale in
  geological materials: A practical introduction to the current
  state-of-the-art}.
\bjtitle{Earth-Science Reviews}
\bvolume{155},
\bfpage{93}--\blpage{128}
(\byear{2016})
\end{barticle}
\endbibitem

\bibitem[\protect\citeauthoryear{Blunt}{2017}]{blunt2017}
\begin{bbook}
\bauthor{\bsnm{Blunt}, \binits{M.J.}}:
\bbtitle{Multiphase Flow in Permeable Media a Pore Scale Perspective}.
\bpublisher{Cambridge University Press, United Kingdom}, \blocation{???}
(\byear{2017}).
\doiurl{10.1017/9781316145098}
\end{bbook}
\endbibitem

\bibitem[\protect\citeauthoryear{Brun et~al.}{2010}]{brun2010}
\begin{barticle}
\bauthor{\bsnm{Brun}, \binits{F.}},
\bauthor{\bsnm{Mancini}, \binits{L.}},
\bauthor{\bsnm{Kasae}, \binits{P.}},
\bauthor{\bsnm{Favretto}, \binits{S.}},
\bauthor{\bsnm{Dreossi}, \binits{D.}},
\bauthor{\bsnm{Tromba}, \binits{G.}}:
\batitle{Pore3d: A software library for quantitative analysis of porous media}.
\bjtitle{Nuclear Instruments and Methods in Physics Research Section A:
  Accelerators, Spectrometers, Detectors and Associated Equipment}
\bvolume{615}(\bissue{3}),
\bfpage{326}--\blpage{332}
(\byear{2010})
\end{barticle}
\endbibitem

\bibitem[\protect\citeauthoryear{Bryant et~al.}{1996}]{bryant}
\begin{barticle}
\bauthor{\bsnm{Bryant}, \binits{S.}},
\bauthor{\bsnm{Mason}, \binits{G.}},
\bauthor{\bsnm{Mellor}, \binits{D.}}:
\batitle{Quantification of spatial correlation in porous media and its effect
  on mercury porosimetry}.
\bjtitle{Journal of Colloid and Interface Science}
\bvolume{177}(\bissue{1}),
\bfpage{88}--\blpage{100}
(\byear{1996})
\end{barticle}
\endbibitem

\bibitem[\protect\citeauthoryear{Ben-Noah}{2024}]{ben_noah_2024_10633605}
\begin{botherref}
\oauthor{\bsnm{Ben-Noah}, \binits{I.}}:
{MATLAB code for locating critical points in binary images based on "Efficient
  pore space characterization based on the curvature of the distance map"}.
Zenodo
(2024).
\doiurl{10.5281/zenodo.10633605} .
\url{https://doi.org/10.5281/zenodo.10633605}
\end{botherref}
\endbibitem

\bibitem[\protect\citeauthoryear{Ben-Noah et~al.}{2024}]{ben2024network}
\begin{botherref}
\oauthor{\bsnm{Ben-Noah}, \binits{I.}},
\oauthor{\bsnm{Hidalgo}, \binits{J.J.}},
\oauthor{\bsnm{Dentz}, \binits{M.}}:
Pore network models to determine the flow statistics and structural controls
  for single-phase flow in partially saturated porous media.
Advances in Water Resources,
104809
(2024)
\end{botherref}
\endbibitem

\bibitem[\protect\citeauthoryear{Borgefors}{1984}]{borgefors1984distance}
\begin{barticle}
\bauthor{\bsnm{Borgefors}, \binits{G.}}:
\batitle{Distance transformations in arbitrary dimensions}.
\bjtitle{Computer vision, graphics, and image processing}
\bvolume{27}(\bissue{3}),
\bfpage{321}--\blpage{345}
(\byear{1984})
\end{barticle}
\endbibitem

\bibitem[\protect\citeauthoryear{Baldwin et~al.}{1996}]{baldwin1996}
\begin{barticle}
\bauthor{\bsnm{Baldwin}, \binits{C.A.}},
\bauthor{\bsnm{Sederman}, \binits{A.J.}},
\bauthor{\bsnm{Mantle}, \binits{M.D.}},
\bauthor{\bsnm{Alexander}, \binits{P.}},
\bauthor{\bsnm{Gladden}, \binits{L.F.}}:
\batitle{Determination and characterization of the structure of a pore space
  from 3d volume images}.
\bjtitle{Journal of Colloid and Interface Science}
\bvolume{181}(\bissue{1}),
\bfpage{79}--\blpage{92}
(\byear{1996})
\doiurl{10.1006/jcis.1996.0358}
\end{barticle}
\endbibitem

\bibitem[\protect\citeauthoryear{Chen et~al.}{2008}]{chen2008}
\begin{barticle}
\bauthor{\bsnm{Chen}, \binits{K.}},
\bauthor{\bsnm{Wang}, \binits{Y.}},
\bauthor{\bsnm{Yang}, \binits{R.}}:
\batitle{Hessian matrix based saddle point detection for granules segmentation
  in 2d image}.
\bjtitle{Journal of Electronics (China)}
\bvolume{25}(\bissue{6}),
\bfpage{228}--\blpage{236}
(\byear{2008})
\doiurl{10.1007/s11767-008-0038-3}
\end{barticle}
\endbibitem

\bibitem[\protect\citeauthoryear{Dong and Blunt}{2009}]{dong2009}
\begin{barticle}
\bauthor{\bsnm{Dong}, \binits{H.}},
\bauthor{\bsnm{Blunt}, \binits{M.J.}}:
\batitle{Pore-network extraction from micro-computerized-tomography images}.
\bjtitle{Phys. Rev. E}
\bvolume{80},
\bfpage{036307}
(\byear{2009})
\doiurl{10.1103/PhysRevE.80.036307}
\end{barticle}
\endbibitem

\bibitem[\protect\citeauthoryear{Eberly et~al.}{1994}]{eberly1994}
\begin{barticle}
\bauthor{\bsnm{Eberly}, \binits{D.}},
\bauthor{\bsnm{Gardner}, \binits{R.}},
\bauthor{\bsnm{Morse}, \binits{B.}},
\bauthor{\bsnm{Pizer}, \binits{S.}},
\bauthor{\bsnm{Scharlach}, \binits{C.}}:
\batitle{Ridges for image analysis}.
\bjtitle{Journal of Mathematical Imaging and Vision}
\bvolume{4},
\bfpage{353}--\blpage{373}
(\byear{1994})
\doiurl{10.1007/BF01262402}
\end{barticle}
\endbibitem

\bibitem[\protect\citeauthoryear{Finney}{2016}]{Finney}
\begin{botherref}
\oauthor{\bsnm{Finney}, \binits{J.}}:
Finney Packing of Spheres.
Digital Rocks Portal
(2016).
\doiurl{10.17612/P78G69}
\end{botherref}
\endbibitem

\bibitem[\protect\citeauthoryear{Gostick}{2017}]{gostick2017}
\begin{barticle}
\bauthor{\bsnm{Gostick}, \binits{J.T.}}:
\batitle{Versatile and efficient pore network extraction method using
  marker-based watershed segmentation}.
\bjtitle{Phys. Rev. E}
\bvolume{96},
\bfpage{023307}
(\byear{2017})
\doiurl{10.1103/PhysRevE.96.023307}
\end{barticle}
\endbibitem

\bibitem[\protect\citeauthoryear{Hoogland et~al.}{2016}]{hoogland2016drainage}
\begin{barticle}
\bauthor{\bsnm{Hoogland}, \binits{F.}},
\bauthor{\bsnm{Lehmann}, \binits{P.}},
\bauthor{\bsnm{Mokso}, \binits{R.}},
\bauthor{\bsnm{Or}, \binits{D.}}:
\batitle{Drainage mechanisms in porous media: From piston-like invasion to
  formation of corner flow networks}.
\bjtitle{Water Resources Research}
\bvolume{52}(\bissue{11}),
\bfpage{8413}--\blpage{8436}
(\byear{2016})
\end{barticle}
\endbibitem

\bibitem[\protect\citeauthoryear{Jim{\'e}nez-Mart{\'\i}nez
  et~al.}{2017}]{jm2017}
\begin{barticle}
\bauthor{\bsnm{Jim{\'e}nez-Mart{\'\i}nez}, \binits{J.}},
\bauthor{\bsnm{Le~Borgne}, \binits{T.}},
\bauthor{\bsnm{Tabuteau}, \binits{H.}},
\bauthor{\bsnm{M{\'e}heust}, \binits{Y.}}:
\batitle{Impact of saturation on dispersion and mixing in porous media:
  Photobleaching pulse injection experiments and shear-enhanced mixing model}.
\bjtitle{Water Resources Research}
\bvolume{53}(\bissue{2}),
\bfpage{1457}--\blpage{1472}
(\byear{2017})
\end{barticle}
\endbibitem

\bibitem[\protect\citeauthoryear{Jiang et~al.}{2007}]{jiang2007}
\begin{botherref}
\oauthor{\bsnm{Jiang}, \binits{Z.}},
\oauthor{\bsnm{Wu}, \binits{K.}},
\oauthor{\bsnm{Couples}, \binits{G.}},
\oauthor{\bsnm{Dijke}, \binits{M.I.J.}},
\oauthor{\bsnm{Sorbie}, \binits{K.S.}},
\oauthor{\bsnm{Ma}, \binits{J.}}:
Efficient extraction of networks from three-dimensional porous media.
Water Resources Research
\textbf{43}(12)
(2007)
\doiurl{10.1029/2006WR005780}
\end{botherref}
\endbibitem

\bibitem[\protect\citeauthoryear{Kaestner et~al.}{2008}]{kaestner2008}
\begin{barticle}
\bauthor{\bsnm{Kaestner}, \binits{A.}},
\bauthor{\bsnm{Lehmann}, \binits{E.}},
\bauthor{\bsnm{Stampanoni}, \binits{M.}}:
\batitle{Imaging and image processing in porous media research}.
\bjtitle{Advances in Water Resources}
\bvolume{31}(\bissue{9}),
\bfpage{1174}--\blpage{1187}
(\byear{2008})
\end{barticle}
\endbibitem

\bibitem[\protect\citeauthoryear{Lin and Cohen}{1982}]{lin1982}
\begin{barticle}
\bauthor{\bsnm{Lin}, \binits{C.}},
\bauthor{\bsnm{Cohen}, \binits{M.}}:
\batitle{Quantitative methods for microgeometric modeling}.
\bjtitle{Journal of Applied Physics}
\bvolume{53}(\bissue{6}),
\bfpage{4152}--\blpage{4165}
(\byear{1982})
\end{barticle}
\endbibitem

\bibitem[\protect\citeauthoryear{Lakemond et~al.}{2011}]{lakemond2011}
\begin{bchapter}
\bauthor{\bsnm{Lakemond}, \binits{R.}},
\bauthor{\bsnm{Fookes}, \binits{C.}},
\bauthor{\bsnm{Sridharan}, \binits{S.}}:
\bctitle{Negative determinant of hessian features}.
In: \bbtitle{2011 International Conference on Digital Image Computing:
  Techniques and Applications},
pp. \bfpage{530}--\blpage{535}
(\byear{2011}).
\bcomment{IEEE}
\end{bchapter}
\endbibitem

\bibitem[\protect\citeauthoryear{Liang et~al.}{2000}]{liang2000}
\begin{barticle}
\bauthor{\bsnm{Liang}, \binits{Z.}},
\bauthor{\bsnm{Ioannidis}, \binits{M.A.}},
\bauthor{\bsnm{Chatzis}, \binits{I.}}:
\batitle{Geometric and topological analysis of three-dimensional porous media:
  Pore space partitioning based on morphological skeletonization}.
\bjtitle{Journal of Colloid and Interface Science}
\bvolume{221}(\bissue{1}),
\bfpage{13}--\blpage{24}
(\byear{2000})
\doiurl{10.1006/jcis.1999.6559}
\end{barticle}
\endbibitem

\bibitem[\protect\citeauthoryear{Lindeberg}{1999}]{lindeberg1999}
\begin{bchapter}
\bauthor{\bsnm{Lindeberg}, \binits{T.}}:
\bctitle{Principles for automatic scale selection}.
In: \beditor{\bsnm{Jähne}, \binits{B.}} (ed.)
\bbtitle{Handbook on Computer Vision and Application},
vol. \bseriesno{{II}},
pp. \bfpage{239}--\blpage{174}.
\bpublisher{Academic Press}, \blocation{???}
(\byear{1999})
\end{bchapter}
\endbibitem

\bibitem[\protect\citeauthoryear{Lee et~al.}{1994}]{lee1994}
\begin{barticle}
\bauthor{\bsnm{Lee}, \binits{T.C.}},
\bauthor{\bsnm{Kashyap}, \binits{R.L.}},
\bauthor{\bsnm{Chu}, \binits{C.N.}}:
\batitle{Building skeleton models via 3-d medial surface axis thinning
  algorithms}.
\bjtitle{CVGIP: Graphical Models and Image Processing}
\bvolume{56}(\bissue{6}),
\bfpage{462}--\blpage{478}
(\byear{1994})
\doiurl{10.1006/cgip.1994.1042}
\end{barticle}
\endbibitem

\bibitem[\protect\citeauthoryear{Lindquist et~al.}{1996}]{lindquist1996}
\begin{barticle}
\bauthor{\bsnm{Lindquist}, \binits{W.B.}},
\bauthor{\bsnm{Lee}, \binits{S.-M.}},
\bauthor{\bsnm{Coker}, \binits{D.A.}},
\bauthor{\bsnm{Jones}, \binits{K.W.}},
\bauthor{\bsnm{Spanne}, \binits{P.}}:
\batitle{Medial axis analysis of void structure in three-dimensional
  tomographic images of porous media}.
\bjtitle{Journal of Geophysical Research: Solid Earth}
\bvolume{101}(\bissue{B4}),
\bfpage{8297}--\blpage{8310}
(\byear{1996})
\doiurl{10.1029/95JB03039}
\end{barticle}
\endbibitem

\bibitem[\protect\citeauthoryear{Lindquist and
  Venkatarangan}{1999}]{lindquist1999}
\begin{barticle}
\bauthor{\bsnm{Lindquist}, \binits{W.B.}},
\bauthor{\bsnm{Venkatarangan}, \binits{A.}}:
\batitle{Investigating 3d geometry of porous media from high resolution
  images}.
\bjtitle{Physics and Chemistry of the Earth, Part A: Solid Earth and Geodesy}
\bvolume{24}(\bissue{7}),
\bfpage{593}--\blpage{599}
(\byear{1999})
\doiurl{10.1016/S1464-1895(99)00085-X}
\end{barticle}
\endbibitem

\bibitem[\protect\citeauthoryear{Lindquist et~al.}{2000}]{lindquist2000}
\begin{barticle}
\bauthor{\bsnm{Lindquist}, \binits{W.B.}},
\bauthor{\bsnm{Venkatarangan}, \binits{A.}},
\bauthor{\bsnm{Dunsmuir}, \binits{J.}},
\bauthor{\bsnm{Wong}, \binits{T.-f.}}:
\batitle{Pore and throat size distributions measured from synchrotron x-ray
  tomographic images of fontainebleau sandstones}.
\bjtitle{Journal of Geophysical Research: Solid Earth}
\bvolume{105}(\bissue{B9}),
\bfpage{21509}--\blpage{21527}
(\byear{2000})
\doiurl{10.1029/2000JB900208}
\end{barticle}
\endbibitem

\bibitem[\protect\citeauthoryear{Mellor}{1989}]{mellor}
\begin{bbook}
\bauthor{\bsnm{Mellor}, \binits{D.W.}}:
\bbtitle{Random Close Packing (RCP) of Equal Spheres: Structure and
  Implications for Use as a Model Porous Medium}.
\bpublisher{Open University (United Kingdom)}, \blocation{???}
(\byear{1989})
\end{bbook}
\endbibitem

\bibitem[\protect\citeauthoryear{Meyer}{1994}]{meyer1994}
\begin{barticle}
\bauthor{\bsnm{Meyer}, \binits{F.}}:
\batitle{Topographic distance and watershed lines}.
\bjtitle{Signal processing}
\bvolume{38}(\bissue{1}),
\bfpage{113}--\blpage{125}
(\byear{1994})
\end{barticle}
\endbibitem

\bibitem[\protect\citeauthoryear{Mason and Mellor}{1991}]{mason}
\begin{bchapter}
\bauthor{\bsnm{Mason}, \binits{G.}},
\bauthor{\bsnm{Mellor}, \binits{D.W.}}:
\bctitle{Analysis of the percolation properties of a real porous material}.
In: \bbtitle{Studies in Surface Science and Catalysis}
vol. \bseriesno{62},
pp. \bfpage{41}--\blpage{50}.
\bpublisher{Elsevier}, \blocation{???}
(\byear{1991})
\end{bchapter}
\endbibitem

\bibitem[\protect\citeauthoryear{Mohammadmoradi}{2017}]{Carbonate}
\begin{botherref}
\oauthor{\bsnm{Mohammadmoradi}, \binits{P.}}:
A Micro CT image of Tight Carbonate.
Digital Rocks Portal
(2017).
\doiurl{10.17612/P77Q2W} .
\url{https://www.digitalrocksportal.org/projects/118}
\end{botherref}
\endbibitem

\bibitem[\protect\citeauthoryear{Mikolajczyk and
  Schmid}{2002}]{mikolajczyk2002}
\begin{bchapter}
\bauthor{\bsnm{Mikolajczyk}, \binits{K.}},
\bauthor{\bsnm{Schmid}, \binits{C.}}:
\bctitle{An affine invariant interest point detector}.
In: \bbtitle{Computer Vision—ECCV 2002: 7th European Conference on Computer
  Vision Copenhagen, Denmark, May 28--31, 2002 Proceedings, Part I 7},
pp. \bfpage{128}--\blpage{142}
(\byear{2002}).
\bcomment{Springer}
\end{bchapter}
\endbibitem

\bibitem[\protect\citeauthoryear{Mehmani and Tchelepi}{2017}]{Mehmani2017}
\begin{barticle}
\bauthor{\bsnm{Mehmani}, \binits{Y.}},
\bauthor{\bsnm{Tchelepi}, \binits{H.A.}}:
\batitle{Minimum requirements for predictive pore-network modeling of solute
  transport in micromodels}.
\bjtitle{Advances in Water Resources}
\bvolume{108},
\bfpage{83}--\blpage{98}
(\byear{2017})
\doiurl{10.1016/j.advwatres.2017.07.014}
\end{barticle}
\endbibitem

\bibitem[\protect\citeauthoryear{Mehmani et~al.}{2020}]{Mehmani2020}
\begin{barticle}
\bauthor{\bsnm{Mehmani}, \binits{A.}},
\bauthor{\bsnm{Verma}, \binits{R.}},
\bauthor{\bsnm{Prodanovi{\'c}}, \binits{M.}}:
\batitle{Pore-scale modeling of carbonates}.
\bjtitle{Marine and Petroleum Geology}
\bvolume{114},
\bfpage{104141}
(\byear{2020})
\end{barticle}
\endbibitem

\bibitem[\protect\citeauthoryear{Neumann et~al.}{2020}]{berea}
\begin{botherref}
\oauthor{\bsnm{Neumann}, \binits{R.}},
\oauthor{\bsnm{Andreeta}, \binits{M.}},
\oauthor{\bsnm{Lucas-Oliveira}, \binits{E.}}:
11 Sandstones: raw, filtered and segmented data.
Digital Rocks Portal
(2020).
\doiurl{10.17612/f4h1-w124} .
\url{http://www.digitalrocksportal.org/projects/317}
\end{botherref}
\endbibitem

\bibitem[\protect\citeauthoryear{Ngom et~al.}{2011}]{ngom2011}
\begin{barticle}
\bauthor{\bsnm{Ngom}, \binits{N.F.}},
\bauthor{\bsnm{Garnier}, \binits{P.}},
\bauthor{\bsnm{Monga}, \binits{O.}},
\bauthor{\bsnm{Peth}, \binits{S.}}:
\batitle{Extraction of three-dimensional soil pore space from microtomography
  images using a geometrical approach}.
\bjtitle{Geoderma}
\bvolume{163}(\bissue{1}),
\bfpage{127}--\blpage{134}
(\byear{2011})
\doiurl{10.1016/j.geoderma.2011.04.013}
\end{barticle}
\endbibitem

\bibitem[\protect\citeauthoryear{Pentland et~al.}{2011}]{pentland2011}
\begin{botherref}
\oauthor{\bsnm{Pentland}, \binits{C.H.}},
\oauthor{\bsnm{El-Maghraby}, \binits{R.}},
\oauthor{\bsnm{Iglauer}, \binits{S.}},
\oauthor{\bsnm{Blunt}, \binits{M.J.}}:
Measurements of the capillary trapping of super-critical carbon dioxide in
  berea sandstone.
Geophysical Research Letters
\textbf{38}(6)
(2011)
\doiurl{10.1029/2011GL046683}
\end{botherref}
\endbibitem

\bibitem[\protect\citeauthoryear{Prodanović et~al.}{2006}]{Prodanovic2006}
\begin{barticle}
\bauthor{\bsnm{Prodanović}, \binits{M.}},
\bauthor{\bsnm{Lindquist}, \binits{W.B.}},
\bauthor{\bsnm{Seright}, \binits{R.S.}}:
\batitle{Porous structure and fluid partitioning in polyethylene cores from 3d
  x-ray microtomographic imaging}.
\bjtitle{Journal of Colloid and Interface Science}
\bvolume{298}(\bissue{1}),
\bfpage{282}--\blpage{297}
(\byear{2006})
\doiurl{10.1016/j.jcis.2005.11.053}
\end{barticle}
\endbibitem

\bibitem[\protect\citeauthoryear{Prodanović et~al.}{2007}]{Prodanovic2007}
\begin{barticle}
\bauthor{\bsnm{Prodanović}, \binits{M.}},
\bauthor{\bsnm{Lindquist}, \binits{W.B.}},
\bauthor{\bsnm{Seright}, \binits{R.S.}}:
\batitle{3d image-based characterization of fluid displacement in a berea
  core}.
\bjtitle{Advances in Water Resources}
\bvolume{30}(\bissue{2}),
\bfpage{214}--\blpage{226}
(\byear{2007})
\doiurl{10.1016/j.advwatres.2005.05.015} .
\bcomment{Pore-scale Modeling}
\end{barticle}
\endbibitem

\bibitem[\protect\citeauthoryear{Pauli et~al.}{1997}]{pauli1997}
\begin{barticle}
\bauthor{\bsnm{Pauli}, \binits{J.}},
\bauthor{\bsnm{Scheying}, \binits{G.}},
\bauthor{\bsnm{Mügge}, \binits{C.}},
\bauthor{\bsnm{Zschunke}, \binits{A.}},
\bauthor{\bsnm{Lorenz}, \binits{P.}}:
\batitle{Determination of the pore widths of highly porous materials with nmr
  microscopy}.
\bjtitle{Fresenius' Journal of Analytical Chemistry}
\bvolume{357}(\bissue{5}),
\bfpage{508}--\blpage{513}
(\byear{1997})
\doiurl{10.1007/s002160050203}
\end{barticle}
\endbibitem

\bibitem[\protect\citeauthoryear{Rabbani et~al.}{2014}]{rabbani2014}
\begin{barticle}
\bauthor{\bsnm{Rabbani}, \binits{A.}},
\bauthor{\bsnm{Jamshidi}, \binits{S.}},
\bauthor{\bsnm{Salehi}, \binits{S.}}:
\batitle{An automated simple algorithm for realistic pore network extraction
  from micro-tomography images}.
\bjtitle{Journal of Petroleum Science and Engineering}
\bvolume{123},
\bfpage{164}--\blpage{171}
(\byear{2014})
\doiurl{10.1016/j.petrol.2014.08.020}
\end{barticle}
\endbibitem

\bibitem[\protect\citeauthoryear{Soille and Ansoult}{1990}]{soille1990}
\begin{barticle}
\bauthor{\bsnm{Soille}, \binits{P.J.}},
\bauthor{\bsnm{Ansoult}, \binits{M.M.}}:
\batitle{Automated basin delineation from digital elevation models using
  mathematical morphology}.
\bjtitle{Signal Processing}
\bvolume{20}(\bissue{2}),
\bfpage{171}--\blpage{182}
(\byear{1990})
\end{barticle}
\endbibitem

\bibitem[\protect\citeauthoryear{Sahimi}{2011}]{sahimi2011}
\begin{bbook}
\bauthor{\bsnm{Sahimi}, \binits{M.}}:
\bbtitle{Flow and Transport in Porous Media and Fractured Rock: from Classical
  Methods to Modern Approaches}.
\bpublisher{John Wiley \& Sons}, \blocation{???}
(\byear{2011})
\end{bbook}
\endbibitem

\bibitem[\protect\citeauthoryear{Solomon and
  Breckon}{2011}]{solomon2011fundamentals}
\begin{bbook}
\bauthor{\bsnm{Solomon}, \binits{C.}},
\bauthor{\bsnm{Breckon}, \binits{T.}}:
\bbtitle{Fundamentals of Digital Image Processing: A Practical Approach with
  Examples in Matlab}.
\bpublisher{John Wiley \& Sons}, \blocation{???}
(\byear{2011})
\end{bbook}
\endbibitem

\bibitem[\protect\citeauthoryear{Safari et~al.}{2021}]{safari2021}
\begin{barticle}
\bauthor{\bsnm{Safari}, \binits{H.}},
\bauthor{\bsnm{Balcom}, \binits{B.J.}},
\bauthor{\bsnm{Afrough}, \binits{A.}}:
\batitle{Characterization of pore and grain size distributions in porous
  geological samples – an image processing workflow}.
\bjtitle{Computers \& Geosciences}
\bvolume{156},
\bfpage{104895}
(\byear{2021})
\doiurl{10.1016/j.cageo.2021.104895}
\end{barticle}
\endbibitem

\bibitem[\protect\citeauthoryear{Smith et~al.}{1987}]{smith1987mercury}
\begin{barticle}
\bauthor{\bsnm{Smith}, \binits{D.}},
\bauthor{\bsnm{Gallegos}, \binits{D.}},
\bauthor{\bsnm{Stermer}, \binits{D.}}:
\batitle{Mercury porosimetry in random sphere packings: Breakthrough pressure
  and pore structure determination}.
\bjtitle{Powder Technology}
\bvolume{53}(\bissue{1}),
\bfpage{11}--\blpage{22}
(\byear{1987})
\end{barticle}
\endbibitem

\bibitem[\protect\citeauthoryear{Silin and Patzek}{2006}]{silin2006}
\begin{barticle}
\bauthor{\bsnm{Silin}, \binits{D.}},
\bauthor{\bsnm{Patzek}, \binits{T.}}:
\batitle{Pore space morphology analysis using maximal inscribed spheres}.
\bjtitle{Physica A: Statistical Mechanics and its Applications}
\bvolume{371}(\bissue{2}),
\bfpage{336}--\blpage{360}
(\byear{2006})
\doiurl{10.1016/j.physa.2006.04.048}
\end{barticle}
\endbibitem

\bibitem[\protect\citeauthoryear{Sheppard et~al.}{2006}]{sheppard2006analysis}
\begin{bchapter}
\bauthor{\bsnm{Sheppard}, \binits{A.}},
\bauthor{\bsnm{Sok}, \binits{R.}},
\bauthor{\bsnm{Averdunk}, \binits{H.}},
\bauthor{\bsnm{Robins}, \binits{V.}},
\bauthor{\bsnm{Ghous}, \binits{A.}}:
\bctitle{Analysis of rock microstructure using high-resolution x-ray
  tomography}.
In: \bbtitle{Proceedings of the International Symposium of the Society of Core
  Analysts},
pp. \bfpage{1}--\blpage{12}
(\byear{2006}).
\bcomment{The Society of Core Analysts Dublin, Ireland}
\end{bchapter}
\endbibitem

\bibitem[\protect\citeauthoryear{Schlueter et~al.}{1997}]{schlueter1997}
\begin{barticle}
\bauthor{\bsnm{Schlueter}, \binits{E.M.}},
\bauthor{\bsnm{Zimmerman}, \binits{R.W.}},
\bauthor{\bsnm{Witherspoon}, \binits{P.A.}},
\bauthor{\bsnm{Cook}, \binits{N.G.W.}}:
\batitle{The fractal dimension of pores in sedimentary rocks and its influence
  on permeability}.
\bjtitle{Engineering Geology}
\bvolume{48}(\bissue{3}),
\bfpage{199}--\blpage{215}
(\byear{1997})
\doiurl{10.1016/S0013-7952(97)00043-4} .
\bcomment{Fractals in Engineering Geology}
\end{barticle}
\endbibitem

\bibitem[\protect\citeauthoryear{{The~MathWorks~Inc.}}{2022}]{MATLAB}
\begin{botherref}
\oauthor{\bsnm{{The~MathWorks~Inc.}}}:
MATLAB version: 9.13.0 (R2022b).
The MathWorks Inc.,
Natick, Massachusetts, United States
(2022).
\url{https://www.mathworks.com}
\end{botherref}
\endbibitem

\bibitem[\protect\citeauthoryear{Thompson et~al.}{2008}]{thompson2008}
\begin{barticle}
\bauthor{\bsnm{Thompson}, \binits{K.E.}},
\bauthor{\bsnm{Willson}, \binits{C.S.}},
\bauthor{\bsnm{White}, \binits{C.D.}},
\bauthor{\bsnm{Nyman}, \binits{S.}},
\bauthor{\bsnm{Bhattacharya}, \binits{J.P.}},
\bauthor{\bsnm{Reed}, \binits{A.H.}}:
\batitle{{Application of a New Grain-Based Reconstruction Algorithm to
  Microtomography Images for Quantitative Characterization and Flow Modeling}}.
\bjtitle{SPE Journal}
\bvolume{13}(\bissue{02}),
\bfpage{164}--\blpage{176}
(\byear{2008})
\doiurl{10.2118/95887-PA}
\end{barticle}
\endbibitem

\bibitem[\protect\citeauthoryear{Warner et~al.}{1989}]{warner1989}
\begin{barticle}
\bauthor{\bsnm{Warner}, \binits{G.S.}},
\bauthor{\bsnm{Nieber}, \binits{J.L.}},
\bauthor{\bsnm{Moore}, \binits{I.D.}},
\bauthor{\bsnm{Geise}, \binits{R.A.}}:
\batitle{Characterizing macropores in soil by computed tomography}.
\bjtitle{Soil Science Society of America Journal}
\bvolume{53}(\bissue{3}),
\bfpage{653}--\blpage{660}
(\byear{1989})
\doiurl{10.2136/sssaj1989.03615995005300030001x}
\end{barticle}
\endbibitem

\bibitem[\protect\citeauthoryear{Wildenschild and
  Sheppard}{2013}]{wildenschild2013}
\begin{barticle}
\bauthor{\bsnm{Wildenschild}, \binits{D.}},
\bauthor{\bsnm{Sheppard}, \binits{A.P.}}:
\batitle{X-ray imaging and analysis techniques for quantifying pore-scale
  structure and processes in subsurface porous medium systems}.
\bjtitle{Advances in Water Resources}
\bvolume{51},
\bfpage{217}--\blpage{246}
(\byear{2013})
\doiurl{10.1016/j.advwatres.2012.07.018} .
\bcomment{35th Year Anniversary Issue}
\end{barticle}
\endbibitem

\bibitem[\protect\citeauthoryear{Yanuka et~al.}{1986}]{yanuka1986}
\begin{barticle}
\bauthor{\bsnm{Yanuka}, \binits{M.}},
\bauthor{\bsnm{Dullien}, \binits{F.A.L.}},
\bauthor{\bsnm{Elrick}, \binits{D.E.}}:
\batitle{Percolation processes and porous media: I. geometrical and topological
  model of porous media using a three-dimensional joint pore size
  distribution}.
\bjtitle{Journal of Colloid and Interface Science}
\bvolume{112}(\bissue{1}),
\bfpage{24}--\blpage{41}
(\byear{1986})
\doiurl{10.1016/0021-9797(86)90066-4}
\end{barticle}
\endbibitem

\bibitem[\protect\citeauthoryear{Youssef et~al.}{2007}]{youssef2007}
\begin{botherref}
\oauthor{\bsnm{Youssef}, \binits{S.}},
\oauthor{\bsnm{Rosenberg}, \binits{E.}},
\oauthor{\bsnm{Gland}, \binits{N.}},
\oauthor{\bsnm{Bekri}, \binits{S.}},
\oauthor{\bsnm{Vizika}, \binits{O.}}:
Quantitative 3d characterisation of the pore space of real rocks: improved
  $\mu$-ct resolution and pore extraction methodology.
Int. Sym. of the Society of Core Analysts
(2007)
\end{botherref}
\endbibitem

\bibitem[\protect\citeauthoryear{Zhao et~al.}{2020}]{zhao2020}
\begin{barticle}
\bauthor{\bsnm{Zhao}, \binits{J.}},
\bauthor{\bsnm{Qin}, \binits{F.}},
\bauthor{\bsnm{Derome}, \binits{D.}},
\bauthor{\bsnm{Kang}, \binits{Q.}},
\bauthor{\bsnm{Carmeliet}, \binits{J.}}:
\batitle{Improved pore network models to simulate single-phase flow in porous
  media by coupling with lattice boltzmann method}.
\bjtitle{Advances in Water Resources}
\bvolume{145},
\bfpage{103738}
(\byear{2020})
\doiurl{10.1016/j.advwatres.2020.103738}
\end{barticle}
\endbibitem

\bibitem[\protect\citeauthoryear{Øren and Bakke}{2002}]{oren2002}
\begin{barticle}
\bauthor{\bsnm{Øren}, \binits{P.-E.}},
\bauthor{\bsnm{Bakke}, \binits{S.}}:
\batitle{Process based reconstruction of sandstones and prediction of transport
  properties}.
\bjtitle{Transport in Porous Media}
\bvolume{46}(\bissue{2}),
\bfpage{311}--\blpage{343}
(\byear{2002})
\doiurl{10.1023/A:1015031122338}
\end{barticle}
\endbibitem

\end{thebibliography}

\end{document}